\documentclass[prl,twocolumn,aps]{revtex4-2}
\usepackage{graphicx}
\usepackage{type1cm}
\usepackage[usenames,svgnames]{xcolor}
\usepackage{color}
\usepackage{url}
\usepackage{natbib}
\usepackage[colorlinks=true,linkcolor=blue,citecolor=blue,breaklinks=true]{hyperref}
\usepackage{hyperref}

\usepackage{amsmath,amssymb,bm,amsthm}
\usepackage{newtxtext,newtxmath}

\DeclareMathOperator{\Tr}{\mathrm{Tr}}
\DeclareMathOperator{\re}{\mathrm{Re}}
\DeclareMathOperator{\argmax}{\mathrm{argmax}}
\newcommand{\del}{\partial}
\usepackage{braket}

\newcommand{\sectionprl}[1]{{\par\it #1.---}}

\begin{document}
\nocite{apsrev41control}

\title{Resolving Discrepancy between Liouvillian Gap and Relaxation Time in Boundary-Dissipated Quantum Many-Body Systems}

\author{Takashi Mori}
\email{
takashi.mori.fh@riken.jp}
\affiliation{
RIKEN Center for Emergent Matter Science (CEMS), Wako 351-0198, Japan
}
\author{Tatsuhiko Shirai}
\affiliation{
Department of Computer Science and Communications Engineering, Waseda University, Tokyo 169-8555, Japan
}

\begin{abstract}
The gap of the Liouvillian spectrum gives the asymptotic decay rate of a quantum dissipative system, and therefore its inverse has been identified as the slowest relaxation time.
In contrary to this common belief, we show that the relaxation time due to diffusive transports in a boundary dissipated many-body quantum system is determined not by the gap or low-lying eigenvalues of the Liouvillian but by superexponentially large expansion coefficients for Liouvillian eigenvectors with non-small eigenvalues at an initial state.
This finding resolves an apparent discrepancy reported in the literature between the inverse of the Liouvillian gap and the relaxation time in dissipative many-body quantum systems.
\end{abstract}
\maketitle

\sectionprl{Introduction}
Understanding the nonequilibrium steady state (NESS) and the relaxation dynamics towards it in a macroscopic open quantum system driven at boundaries is a central problem of nonequilibrium statistical physics and condensed matter physics~\cite{Dhar2008,Prosen2008,Prosen2011,Znidaric2015}.
This problem is of practical importance in the context of quantum technologies since recent advance in experiments using ultra-cold atoms allows us to implement highly controllable dissipative dynamics~\cite{Barreiro2011,Barontini2013,Tomita2017}.

Since the relaxation to the NESS takes place via the transport of conserved quantities, its timescale is determined by the transport property of the bulk Hamiltonian.
If the slowest process is the diffusive transport, the relaxation time is proportional to $L^2$, where $L$ is the diameter of the system, while if all the transports are ballistic, it is proportional to $L$.

The dynamics of an open quantum system is generated by the Liouvillian superoperator, and thus the inverse of the gap of the Liouvillian spectrum has been identified as the relaxation time.
It is then natural to expect that, in the thermodynamic limit, the Liouvillian gap closes as $L^{-2}$ in a boundary-dissipated quantum chaotic system in which transports are diffusive.
However, numerical results for finite systems by \v{Z}nidari\v{c}~\cite{Znidaric2015} show that the Liouvillian gap closes slower than $L^{-2}$ in various boundary-dissipated systems with diffusive transports.
Is such a large gap of the Liouvillian just a finite-size effect and should the Liouvillian gap always close as $L^{-2}$ for sufficiently large system sizes?

In this Letter, we address the above question.
It turns out that the relaxation time due to diffusive transports is originated not from low-lying eigenvalues of the Liouvillian but from extraordinarily large ($\sim e^{O(L^2)}$) expansion coefficients at an initial state, which is due to non-Hermiticity of the Liouvillian.
Slowly vanishing gap $g^{-1}=o(L^2)$ for large $L$ does not contradict the relaxation time of $O(L^2)$ due to diffusive transports.
Our result is contrary to a common belief that the Liouvillian gap determines the relaxation time, and hence we should take special care for discussing the relaxation time in dissipative quantum systems.

\sectionprl{Liouvillian eigenvalues and eigenvectors}
Under Markov approximation, the dissipative dynamics of the density matrix $\rho(t)$ of an open quantum system is described by the Lindblad equation~\cite{Lindblad1976,Breuer_text}
\begin{equation}
\left\{
\begin{split}
&\frac{d}{dt}\rho(t)=\mathcal{L}\rho(t); \\
&\mathcal{L}\rho=-i[\hat{H},\rho]+\sum_a\left(\hat{L}_a\rho\hat{L}_a^\dagger-\frac{1}{2}\{\hat{L}_a^\dagger\hat{L}_a,\rho\}\right),
\end{split}
\right.
\label{eq:Lindblad}
\end{equation}
where $H$ is the bulk Hamiltonian and $\{\hat{L}_a\}$ are called the Lindblad operators that characterize the dissipation.
The commutator and the anti-commutator are denoted by $[\cdot,\cdot]$ and $\{\cdot,\cdot\}$, respectively.
We consider a one-dimensional lattice system and assume that dissipation acts only at two ends of the system, i.e., Lindblad operators are local operators acting nontrivially to either the left or right boundary.

The superoperator $\mathcal{L}$ is called the Liouvillian.
Its complex eigenvalues are denoted by $\{\lambda_n\}_{n=0,1,2,\dots}$, which are sorted as $0=\lambda_0>\re\lambda_1\geq\re\lambda_2\geq\dots$ (we assume that the zero eigenvalue is not degenerate)~\footnote{Although a non-Hermitian operator may not be diagonalizable, such a situation is believed to be rare. Indeed, the Liouvillian is always diagonalizable in our numerical calculations. We therefore assume the diagonalizability of the Liouvillian in theoretical considerations.}.
The corresponding right and left eigenvectors are denoted by $\{\rho_n\}$ and $\{\pi_n\}$, respectively.
We normalize the eigenvectors using the trace norm, i.e.,
\begin{equation}
\|\rho_n\|_\mathrm{tr}=\|\pi_n\|_\mathrm{tr}=1,
\label{eq:norm}
\end{equation}
where $\|\hat{A}\|_\mathrm{tr}:=\Tr\sqrt{\hat{A}^\dagger\hat{A}}$.
Let us define the inner product of two operators $\hat{A}$ and $\hat{B}$ as $\braket{\hat{A},\hat{B}}=\Tr\hat{A}^\dagger\hat{B}$.
The orthogonality of eigenvectors is then expressed as $\braket{\pi_n,\rho_m}=0$ for all $n\neq m$.
The right eigenvector $\rho_0$ with zero eigenvalue corresponds to the density matrix of the NESS, so we write $\rho_0=\rho_\mathrm{ss}$.
If the initial state $\rho(0)$ is expanded as
\begin{equation}
\rho(0)=\rho_\mathrm{ss}+\sum_{n\neq 0}c_n\rho_n,
\end{equation}
the state at time $t>0$ is given by
\begin{equation}
\rho(t)=\rho_\mathrm{ss}+\sum_{n\neq 0}c_ne^{\lambda_nt}\rho_n.
\end{equation}
The distance $d_T$ between $\rho(t)$ and the NESS is measured by the trace norm~\footnote{We can show that, for any bounded operator $\hat{O}$, $|\Tr\hat{O}\rho(t)-\Tr\hat{O}\rho_\mathrm{ss}|\leq \|\hat{O}\|d_T(t)$, where $\|\hat{O}\|$ denotes the operator norm of $\hat{O}$. A small trace distance $d_T(t)$ ensures a small difference between the expectation values of $\hat{O}$ at $\rho(t)$ and at $\rho_\mathrm{ss}$. The trace norm thus provides us a natural measure of the distance between two density matrices.} as
\begin{equation}
d_T(t)=\|\rho(t)-\rho_\mathrm{ss}\|_\mathrm{tr},
\end{equation}
which monotonically decreases with $t$~\cite{Ruskai1994}.
We define the relaxation time as the time $\tau$ satisfying $d_T(\tau)=2\epsilon$ with a fixed constant $\epsilon\in(0,1)$ (the precise value of $\epsilon$ does not matter in our purpose).

The Liouvillian gap $g$ is defined as
\begin{equation}
g=-\re\lambda_1,
\end{equation}
which determines the asymptotic decay rate~\cite{Kessler2012} and also carries information on some properties of the NESS~\cite{Poulin2010,Kessler2012,Hoening2012,Shirai2020}.
For sufficiently large $t$, $\rho(t)-\rho_\mathrm{ss}\sim e^{-gt}\rho_1$ and it is expected that $\tau\lesssim 1/g$.

\sectionprl{Superexponentially large $c_n$}
\v{Z}nidari\v{c}~\cite{Znidaric2015} numerically showed that $g\propto L^{-z}$ with $1\leq z<2$, although the bulk Hamiltonian is chaotic and there exist diffusive transports, which implies $\tau\propto L^2$.
This result violates the relation $\tau\lesssim 1/g$.

We want to understand how this discrepancy is resolved.
Although the discussion below is general, for clarity we focus on the hard-core Bose-Hubbard model under boundary dephasing dissipation.
The bulk Hamiltonian is given by
\begin{align}
\hat{H}=-h\sum_{i=1}^{L-1}\left(\hat{b}_{i+1}^\dagger\hat{b}_i+\hat{b}_i^\dagger\hat{b}_{i+1}\right)-h'\sum_{i=1}^{L-2}\left(\hat{b}_{i+2}^\dagger\hat{b}_i+\hat{b}_i^\dagger\hat{b}_{i+2}\right)
\nonumber \\
+U\sum_{i=1}^{L-1}\left(\hat{n}_i-\frac{1}{2}\right)\left(\hat{n}_{i+1}-\frac{1}{2}\right)+U'\sum_{i=1}^{L-2}\left(\hat{n}_i-\frac{1}{2}\right)\left(\hat{n}_{i+2}-\frac{1}{2}\right),
\label{eq:BH}
\end{align}
where $\hat{b}_i$, $\hat{b}_i^\dagger$ are annihilation and creation operators of a hard-core boson at site $i$, respectively.
The number operator is denoted by $\hat{n}_i=\hat{b}_i^\dagger\hat{b}_i$.
We fix the parameters as $h=U=1$ and $h'=U'=0.24$.
This model is known to be chaotic~\cite{Santos2010a}.

Dephasing dissipation on the first and the last site corresponds to the Lindblad operators $\{\hat{L}_a\}_{a=1,2}$ with
\begin{equation}
\hat{L}_1=2\hat{b}_1^\dagger\hat{b}_1, \quad \hat{L}_2=2\hat{b}_L^\dagger\hat{b}_L.
\label{eq:dephasing}
\end{equation}
This model conserves the total particle number $N=\sum_{i=1}^L\hat{n}_i$, and hence we restrict ourselves to the sector of $N=L/2$ for $L$ even and $N=(L-1)/2$ for $L$ odd.
In this model, the gap closes as $g\sim L^{-1.6}$.

In Supplementary Material (SM), we also investigate another choice of dissipation~\cite{SM}, but the result presented below is not sensitive to the specific choice of boundary dissipation.

\begin{figure}[t]
\centering
\includegraphics[width=0.95\linewidth]{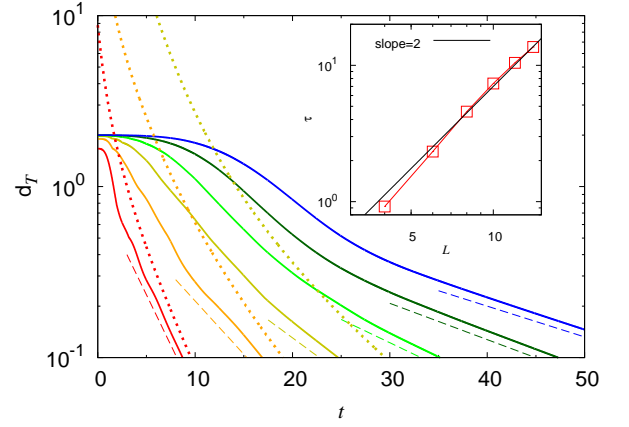}
\caption{Dynamics of the trace distance $d_T$ for various system sizes in the Bose-Hubbard model under boundary dephasing dissipation.
Solid lines show $d_T(t)$ for $L=4, 6, 8, 10, 12, 14$ from left to right.
Dotted lines are upper bounds $\bar{d}_T(t)$ for $L=4, 6, 8$ from left to right.
The dashed lines have the slope $-g$ for each $L$.
Inset shows the relaxation time $\tau$ at which $d_T$ becomes 1.5.
Clearly $\tau\propto L^2$, which is consistent with the timescale of diffusive transports.}
\label{fig:dynamics}
\end{figure}

Figure~\ref{fig:dynamics} shows the actual time evolution of $d_T(t)$ for various system sizes $L$ in the model under the dephasing dissipation.
In the initial state, all the left-half sites are occupied and all the right-half sites are empty, i.e., $n_1=n_2=\dots=n_{\lfloor L/2\rfloor}=1$ and $n_{\lfloor L/2\rfloor+1}=\dots=n_L=0$.
The inverse of the Liouvillian gap is relevant only in the later stage, which is indicated by dashed lines in Fig.~\ref{fig:dynamics}, and the relaxation time $\tau$ does not satisfy $\tau\lesssim 1/g$.
Instead, we find $\tau\propto L^2$, which is expected by the presence of diffusive transports (see the inset of Fig.~\ref{fig:dynamics}).

To understand how diffusive relaxation timescale arises despite $z<2$, let us consider the following upper bound of the trace distance:
\begin{equation}
d_T(t)\leq\sum_{n\neq 0}|c_n|e^{\re(\lambda_n)t}=:\bar{d}_T(t).
\label{eq:upper_bound}
\end{equation}
The upper bound $\bar{d}_T(t)$ is also plotted up to $L=8$ in Fig.~\ref{fig:dynamics}.
We notice that, at $t=0$, $\bar{d}_T(0)\gg 1$ for large $L$.
Since $d_T(0)\leq 2$, the upper bound $\bar{d}_T(t)$ is not tight at all for small $t$.
However, we find that at later times $\bar{d}_T(t)$ captures the decay of $d_T(t)$ and gives a good estimate of the relaxation time; if we define $\tau'$ by $\bar{d}_T(\tau')=2\epsilon$, $\tau'$ shows the same system size dependence as $\tau$.
We can therefore use $\bar{d}_T(t)$ in estimating $\tau$.
In particular, from eq.~(\ref{eq:upper_bound}), $\tau$ is estimated by the condition $|c_n|e^{\re(\lambda_n)\tau}\ll1$ for all $n\neq 0$.

It is an important observation that $\bar{d}_T(0)$ rapidly grows with $L$ as $\bar{d}_T(0)=e^{O(L^2)}$~\cite{SM}.
This behavior implies that some expansion coefficients $c_n$ should be $e^{O(L^2)}$~\footnote{Indeed, we have the inequality $\bar{d}_T(0)=\sum_{n\neq 0}|c_n|\leq D^2\max_{n\neq 0}|c_n|$, where $D$ is the dimension of the Hilbert space.
Because $D=e^{O(L)}$, we obtain $\max_{n\neq 0}|c_n|\geq \bar{d}_T(0)/D^2=e^{O(L^2)}$ whenever $\bar{d}_T(0)=e^{O(L^2)}$.
}.
When such an anomalously large expansion coefficient $c_n$ appears at $|\re(\lambda_n)|=O(1)$, the condition $|c_n|e^{\re(\lambda_n)\tau}\ll 1$ leads to $\tau\sim L^2/|\re(\lambda_n)|\propto L^2$.
In this way, \textit{superexponentially large expansion coefficients give the relaxation time due to diffusive transports}.
In SM, expansion coefficients are explicitly computed for the above initial state, which confirms that superexponentially large expansion coefficients certainly appear~\cite{SM}.

An expansion coefficient is expressed as
\begin{equation}
c_n=\frac{\braket{\pi_n,\rho(0)}}{\braket{\pi_n,\rho_n}}.
\label{eq:coeff}
\end{equation}
Its numerator cannot be large since $|\braket{\pi_n,\rho(0)}|\leq\|\pi_n\|_\mathrm{tr}\|\rho(0)\|_\mathrm{tr}=1$.
Superexponentially large $c_n$ must stem from an anomalously small overlap between the left and the right eigenvectors: $|\braket{\pi_n,\rho_n}|=e^{-O(L^2)}$.

\sectionprl{Evaluation of the relaxation time}
The next problem is to clarify which eigenmode is responsible for diffusive relaxation.
To address it, let us first consider an initial state with a single excited mode
\begin{equation}
\rho(0)=\rho_\mathrm{ss}+c_n\rho_n,
\end{equation}
where we assume that $\lambda_n$ is real for simplicity (in this case $\rho_n=\rho_n^\dagger$ and $\pi_n=\pi_n^\dagger$ hold).
We have $d_T(t)=|c_n|e^{\re(\lambda_n)t}$.
It should be noted that $d_T(t)$ is bounded by 2, and hence $|c_n|$ is restricted by
\begin{equation}
|c_n|\leq 2.
\label{eq:condition_c}
\end{equation}
No large expansion coefficient appears and the relaxation time is thus given by $\tau\sim|\re\lambda_n|^{-1}\leq g^{-1}$.

In this way, for an initial state with a single excited \textit{right} eigenvector, eq.~(\ref{eq:condition_c}) must be satisfied and expansion coefficients cannot grow with $L$.
However, for generic initial states, eq.~(\ref{eq:condition_c}) does not need to hold and $|c_n|$ may take a much larger value.
Since an expansion coefficient is given by eq.~(\ref{eq:coeff}), $n$th mode would be strongly excited by considering an initial state with a single \textit{left} eigenvector excited:
\begin{equation}
\rho(0)=\rho_\mathrm{ss}+a_n\pi_n,
\end{equation}
where we again assume that $\lambda_n$ is real, for simplicity.
Similarly to eq.~(\ref{eq:condition_c}), $a_n$ satisfies
\begin{equation}
|a_n|\leq 2.
\label{eq:condition_a}
\end{equation}
Now let us expand this state in terms of the right eigenvectors, $\rho(0)=\rho_\mathrm{ss}+\sum_{m\neq 0}c_m\rho_m$ with
\begin{equation}
c_m=\frac{\braket{\pi_m,\rho(0)}}{\braket{\pi_m,\rho_m}}=\frac{\braket{\pi_m,\pi_n}}{\braket{\pi_m,\rho_m}}a_n.
\end{equation}
For $m=n$, we have $c_n=a_n\braket{\pi_n,\pi_n}/\braket{\pi_n,\rho_n}$, and thus by using eq.~(\ref{eq:condition_a}), we obtain
\begin{equation}
|c_n|\leq 2\left|\frac{\braket{\pi_n,\pi_n}}{\braket{\pi_n,\rho_n}}\right|=:2\Phi_n.
\label{eq:general_bound}
\end{equation}
Since $\|\pi_n\|_\mathrm{tr}^2/D\leq |\braket{\pi_n,\pi_n}|\leq \|\pi_n\|_\mathrm{tr}^2$ and $\|\pi_n\|_\mathrm{tr}=1$ hold, we have $\Phi_n=e^{O(L^2)}$ whenever $|\braket{\pi_n,\rho_n}|=e^{-O(L^2)}$.
Thus expansion coefficients can be superexponentially large in this class of initial states, and hence we can more precisely study which eigenmode is related to diffusion.

When $\lambda_n$ is not real, we have to consider $\rho(0)=\rho_\mathrm{ss}+a_n\pi_n+a_n^*\pi_n^\dagger$ to ensure the Hermiticity of the density matrix.
In this case, by defining $\Phi_n$ as
\begin{equation}
\Phi_n=\frac{1}{|\braket{\pi_n,\rho_n}|}\max_{\theta\in[0,\pi]}\frac{|\braket{\pi_n,\pi_ne^{i\theta}+\pi_n^\dagger e^{-i\theta}}|}{\|\pi_ne^{i\theta}+\pi_n^\dagger e^{-i\theta}\|_\mathrm{tr}},
\end{equation}
it is shown that $|c_n|\leq 2\Phi_n$~\footnote{In numerical calculations, the maximization with respect to $\theta$ is done by discretizing it into $\theta=\pi/50, 2\pi/50,\dots,49\pi/50$.}.

Since the relaxation time $\tau$ should satisfy the condition $|c_n|e^{\re(\lambda_n)\tau}\leq 2\epsilon$ for some fixed small constant $\epsilon\in(0,1)$, we obtain
\begin{equation}
\tau\sim\frac{\ln\epsilon}{\re\lambda_n}+\frac{\ln\Phi_n}{|\re(\lambda_n)|}.
\label{eq:tau}
\end{equation}
The first term of eq.~(\ref{eq:tau}) gives a contribution to the relaxation time that is roughly bounded from above by $g^{-1}$.
When $z<2$, this contribution is $o(L^2)$, which does not explain the relaxation time due to diffusive transports.
We therefore focus on the second term of eq.~(\ref{eq:tau}),
\begin{equation}
\tau_n:=\frac{\ln\Phi_n}{|\re(\lambda_n)|}.
\label{eq:tau_n}
\end{equation}

If the Liouvillian were Hermitian, $\rho_n=\pi_n$ and $\Phi_n=1$.
Therefore, the divergence of $\ln\Phi_n$ in the thermodynamic limit, which alters the system-size dependence of the relaxation time, is a result of non-Hermiticity of the Liouvillian.

In a recent work~\cite{Haga2020}, it is shown that an exponentially small overlap $\braket{\pi_n,\rho_n}$ such that $\ln\Phi_n\propto L$ arises due to the localization of left and right eigenmodes at the opposite boundaries of the system in a single-particle model under bulk dissipation (see also Ref.~\cite{Song2019}).
As far as we have calculated, however, superexponentially small overlaps $|\braket{\pi_n,\rho_n}|=e^{-O(L^2)}$ in a boundary-dissipated many-body system are not simply explained by such localization.

Below, we numerically show that $\tau_n\propto L^2$ for \textit{typical} $n$.
This result indicates that superexponentially large expansion coefficients appear for generic initial states. 
It also implies that, contrary to a common belief, diffusive transports are not necessarily associated with the gap or low-lying eigenvalues of the Liouvillian.

\begin{figure}
\centering
\includegraphics[width=0.8\linewidth]{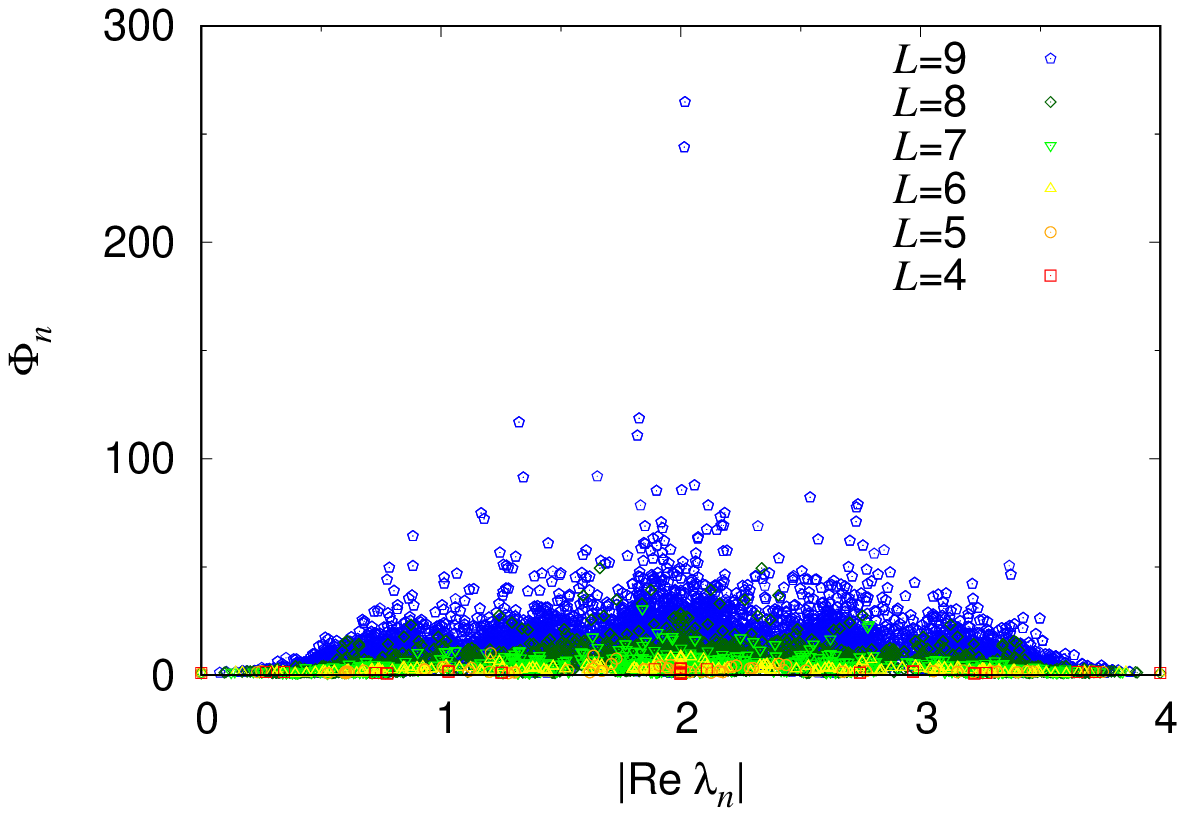}
\includegraphics[width=0.8\linewidth]{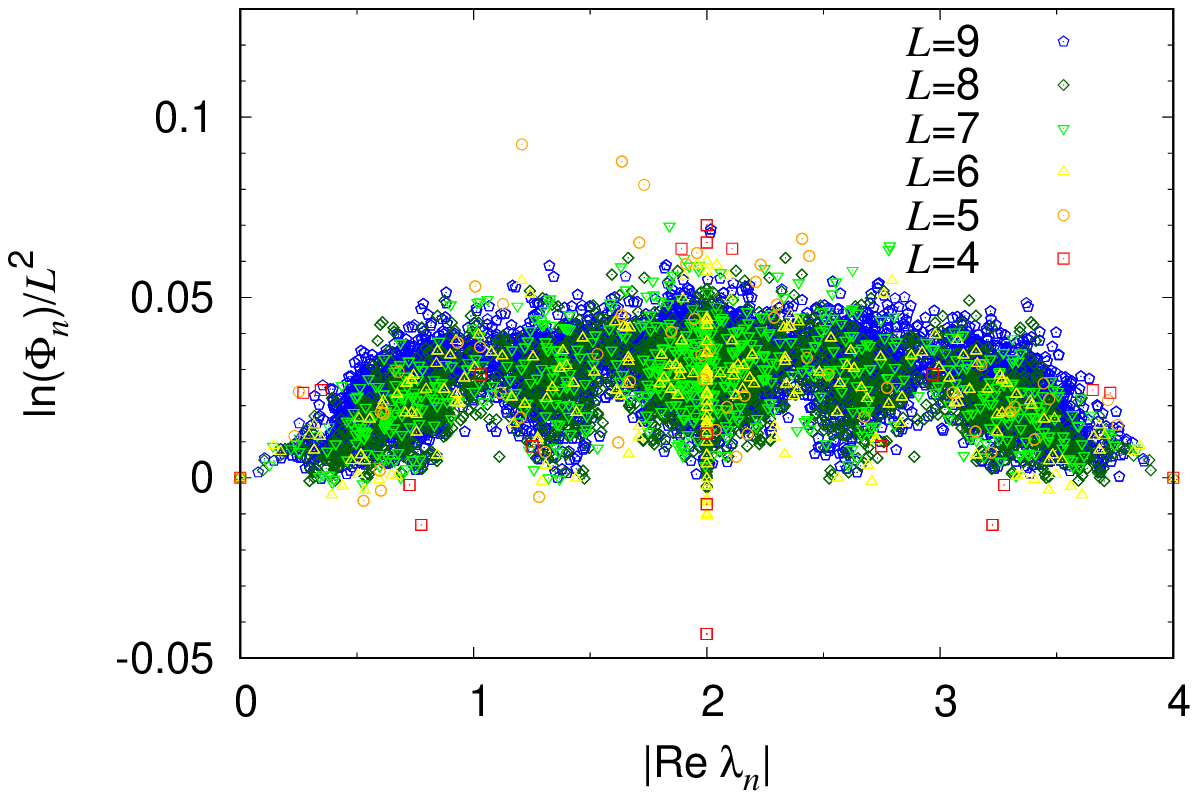}
\caption{Numerically calculated values of $\Phi_n$ (Top) and the rescaled quantity $\ln(\Phi_n)/L^2$ (Bottom).
The horizontal axis is $|\re\lambda_n|$.
After the rescaling, the data for different system sizes collapse, which indicates $\Phi_n=e^{O(L^2)}$ for typical $n$.}
\label{fig:dephasing}
\end{figure}

Now we present numerical results obtained by the exact diagonalization. 
Figure~\ref{fig:dephasing} shows $\Phi_n$ as a function of $|\re\lambda_n|$ up to $L=9$.
We see that the values of $\{\Phi_n\}$ rapidly grow with $L$.
In the same figure, the rescaled quantity $\ln(\Phi_n)/L^2$ is also shown.
The system-size dependence disappears after rescaling, which means that $\Phi_n$ behaves as $\Phi_n=e^{O(L^2)}$ \textit{for typical $n$}.

\begin{figure}
\centering
\includegraphics[width=0.8\linewidth]{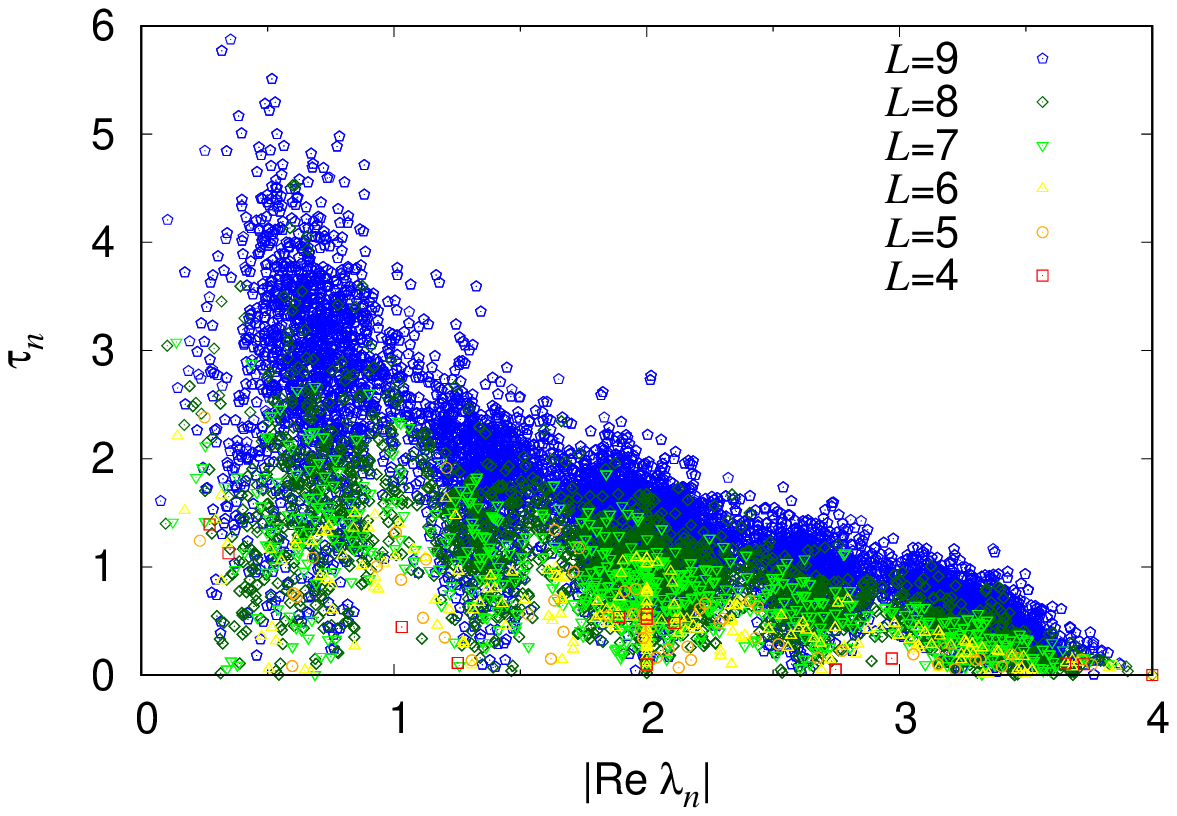}
\includegraphics[width=0.8\linewidth]{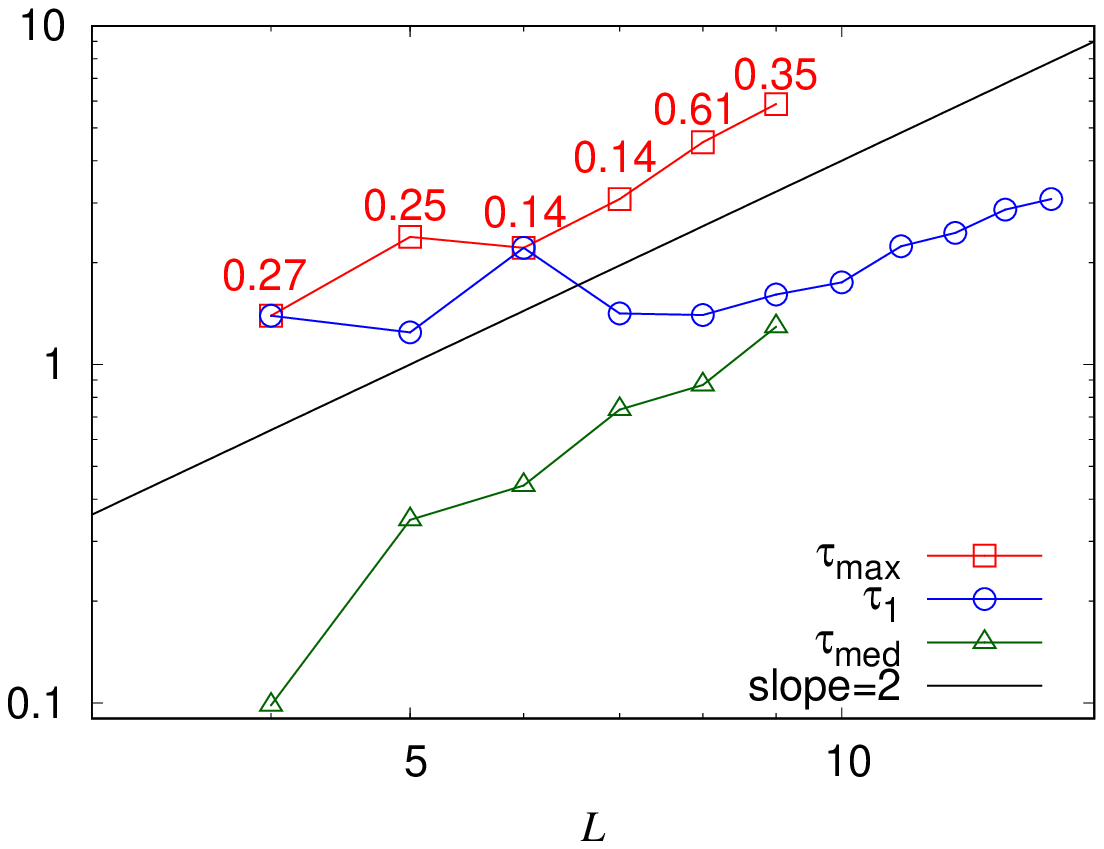}
\caption{(Top) $\tau_n$ for varying system sizes $L$.
(Bottom) Log-log plot of $\tau_\mathrm{max}$, $\tau_\mathrm{med}$, and $\tau_1$ against $L$.
We find $\tau_\mathrm{max},\tau_\mathrm{med}\propto L^2$, which agrees with diffusive transports, while $\tau_1$ looks increasing more slowly with $L$.
The number indicated for each plot point of $\tau_\mathrm{max}$ corresponds to $|\re\lambda_{n^*}|$ with $n^*=\argmax_n\tau_n$.}
\label{fig:dephasing_tau}
\end{figure}

Figure.~\ref{fig:dephasing_tau} shows $\tau_n$ for varying $L$ and the system-size dependence of $\tau_\mathrm{max}$, $\tau_\mathrm{med}$ and $\tau_1$.
Here, $\tau_\mathrm{max}=\max_n\tau_n$ and $\tau_\mathrm{med}$ is the median of $\{\tau_n\}$.
We see that $\tau_\mathrm{max}, \tau_\mathrm{med}\propto L^2$, which means that the relaxation time due to diffusive transports typically appears in the class of initial states $\rho(0)=\rho_\mathrm{ss}+a_n\pi_n$.
On the other hand, $\tau_1$ increases with $L$ but slower than $O(L^2)$, which means that the first excited eigenmode giving the Liouvillian gap does not produce diffusive relaxation.

\sectionprl{Discussion}
We have investigated the gap discrepancy problem which was reported in Ref.~\cite{Znidaric2015}.
This discrepancy is resolved by considering the system-size dependence of expansion coefficients $c_n=e^{O(L^2)}$.
Although it is well known that large expansion coefficients may appear when a given state is expanded by a non-orthogonal basis, our finding is that they have a physical consequence: they influence the system-size dependence of the relaxation time.
We conclude our Letter with some remarks.

Firstly, our theoretical argument is rather generic, but the behavior of $\Phi_n=e^{O(L^2)}$ for typical eigenmodes is closely related to conserved currents induced by boundary dissipation.
In SM, we show that the same thing happens in another boundary-dissipated model, but we obtain qualitatively different behavior of $\Phi_n$ in a bulk-dissipated system without any conserved quantities~\cite{SM}.
In the bulk-dissipated model, we have no superexponentially large $\Phi_n$.
Instead, we have exponentially large $\Phi_n=e^{O(L)}$ for $\re\lambda_n=O(L)$, which indicates that $\tau_n$ does not depend on the system size as expected.
In this way, there is an important difference between boundary- and bulk-dissipated systems, but in both cases, we need to consider the system-size dependence of expansion coefficients for an accurate evaluation of the relaxation time.

Secondly, although it is well known that the overlap of left and right eigenvectors of a non-Hermitian operator can be very small if the corresponding eigenvalue is almost degenerate (i.e. close to an exceptional point), it is hard to understand the behavior $\Phi_n=e^{O(L^2)}$ (or $\braket{\pi_n,\rho_n}=e^{-O(L^2)}$) as such a near-degeneracy effect.
Indeed, the typical eigenvalue distance of the Liouvillian is found to be $e^{-O(L)}$, which is much larger than $e^{-O(L^2)}$~\cite{SM}.

Thirdly, large expansion coefficients $|c_n|=e^{O(L^2)}$ at non-small eigenvalues $|\re\lambda_n|=O(L^0)$ implies that the trace distance to the stationary state is almost constant up to a time $\tau=O(L^2)$ but suddenly decays over a narrow window of time $\Delta t=O(L^0)$ (we can see this behavior in Fig.~\ref{fig:dynamics}).
This is interpreted as a quantum analogue of the \textit{cutoff phenomenon}~\cite{Vernier2020}, which has been studied in classical Markov processes~\cite{Aldous1986, Berestycki2016}.

Finally, our theoretical argument applies to generic non-Hermitian dynamics, and large expansion coefficients can appear in other settings like classical Markov processes~\cite{Derrida1993,Derrida2001,DeGier2005}.
We demonstrated in SM that the boundary-driven symmetric simple exclusion process shows the divergence of $\{\Phi_n\}$, but there are some differences from the quantum model discussed so far~\cite{SM}.
We find that typically $\Phi_n\sim e^{O(L)}$ (not $e^{O(L^2)}$), and the diffusive relaxation time stems from low-lying eigenmodes with $|\re\lambda_n|\lesssim 1/L$ [not from eigenmodes with $|\re\lambda_n|=O(L^0)$].
Thus our work has nontrivial implications beyond the context of quantum dissipative systems and more detailed studies are desired.

\begin{acknowledgments}
We would like to thank Hosho Katsura, Eric Vernier, and Marko \v{Z}nidari\v{c} for useful comments.
This work was supported by Japan Society for the Promotion of Science KAKENHI Grants No. 19K14622 and No. 18K13466.
The numerical calculations have been done mainly on the supercomputer system at Institute for Solid State Physics, University of Tokyo.
\end{acknowledgments}

\bibliography{apsrevcontrol,physics}

\clearpage
\begin{widetext}

\begin{center}
\textbf{\Large Supplemental Material}

\bigskip
Takashi Mori$^{1}$ and Tatsuhiko Shirai$^{2}$\\
\textit{
${}^1$RIKEN Center for Emergent Matter Science (CEMS), Wako 351-0198, Japan
}\\
\textit{
${}^2$Department of Computer Science and Communications Engineering, Waseda University, Tokyo 169-8555, Japan
}

\end{center}

\setcounter{equation}{0}
\def\theequation{S\arabic{equation}}
\setcounter{figure}{0}
\def\thefigure{S\arabic{figure}}

\section{A. Expansion coefficients for a specific initial state}

In the main text, we mentioned that we have superexponentially large expansion coefficients $c_n=e^{O(L^2)}$ for an initial state in which all the left-half sites are occupied and all the right-half sites are empty, i.e., $n_1=n_2=\dots=n_{\lfloor L/2\rfloor}=1$ and $n_{\lfloor L/2\rfloor+1}=\dots=n_L=0$.
Here we explicitly confirm it by numerically computing $\bar{d}_T(0)=\sum_n|c_n|$ and, more directly, expansion coefficients for this kind of initial states.

In Fig.~\ref{fig:expansion} (a), the system-size dependence of $\bar{d}_T(0)$ for this initial state is shown.
We find that $\ln\bar{d}_T(0)\propto L^2$ as is mentioned in the main text.
This is an evidence that some expansion coefficients grow as $e^{O(L^2)}$.

We plot $\{|c_n|\}$ for various $L$ in Fig.~\ref{fig:expansion} (b).
We see explosive growth of some $|c_n|$ as $L$ increases.
In Fig.~\ref{fig:expansion} (c), the scaled quantities $\{\ln(|c_n|)/L^2\}$ are shown.
The relaxation time is estimated as $\tau\sim\max_n[\ln(|c_n|)/|\re\lambda_n|]$, which is plotted in Fig.~\ref{fig:expansion} (d).
We see that it looks consistent with $\tau\propto L^2$, although data for larger system sizes are needed to definitely conclude $\tau\propto L^2$ (or $|c_n|=e^{O(L^2)}$).

\begin{figure}[b]
\centering
\begin{tabular}{cc}
(a) & (b) \\
\includegraphics[width=0.4\linewidth]{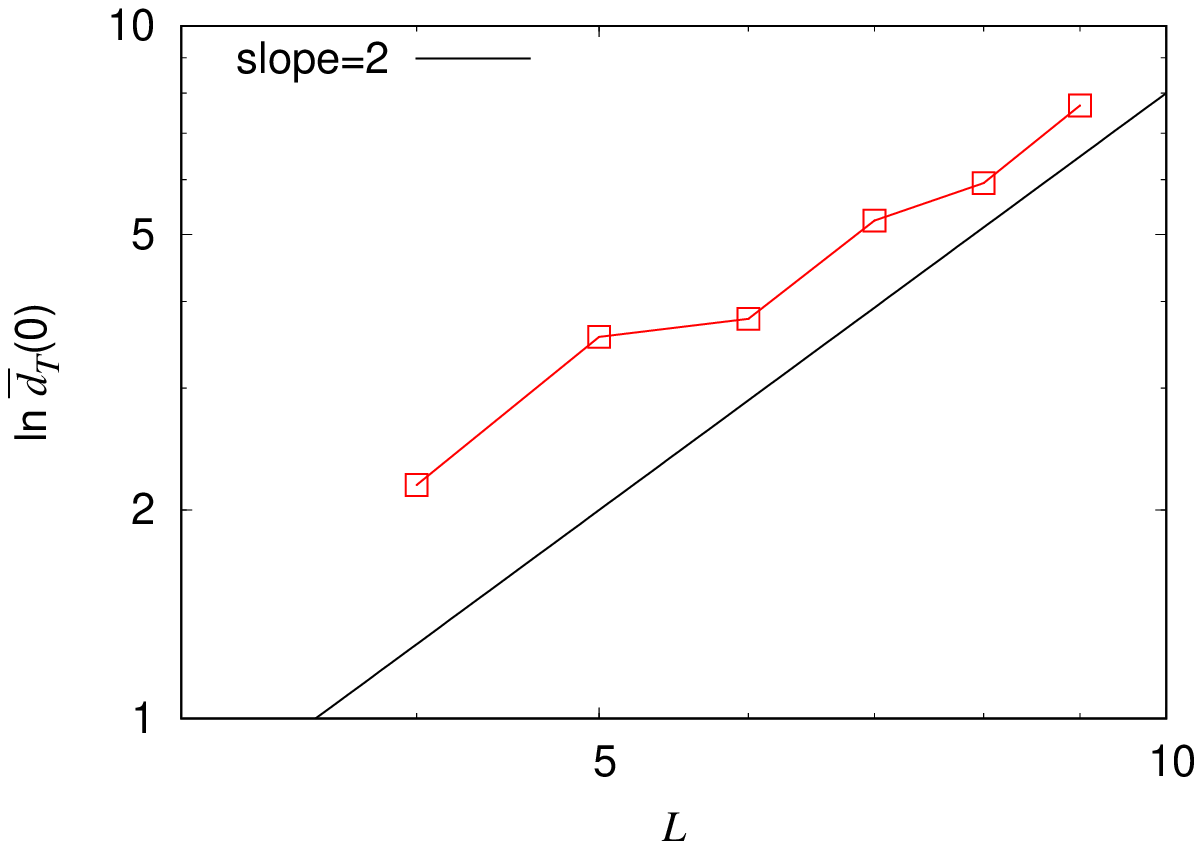}&
\includegraphics[width=0.4\linewidth]{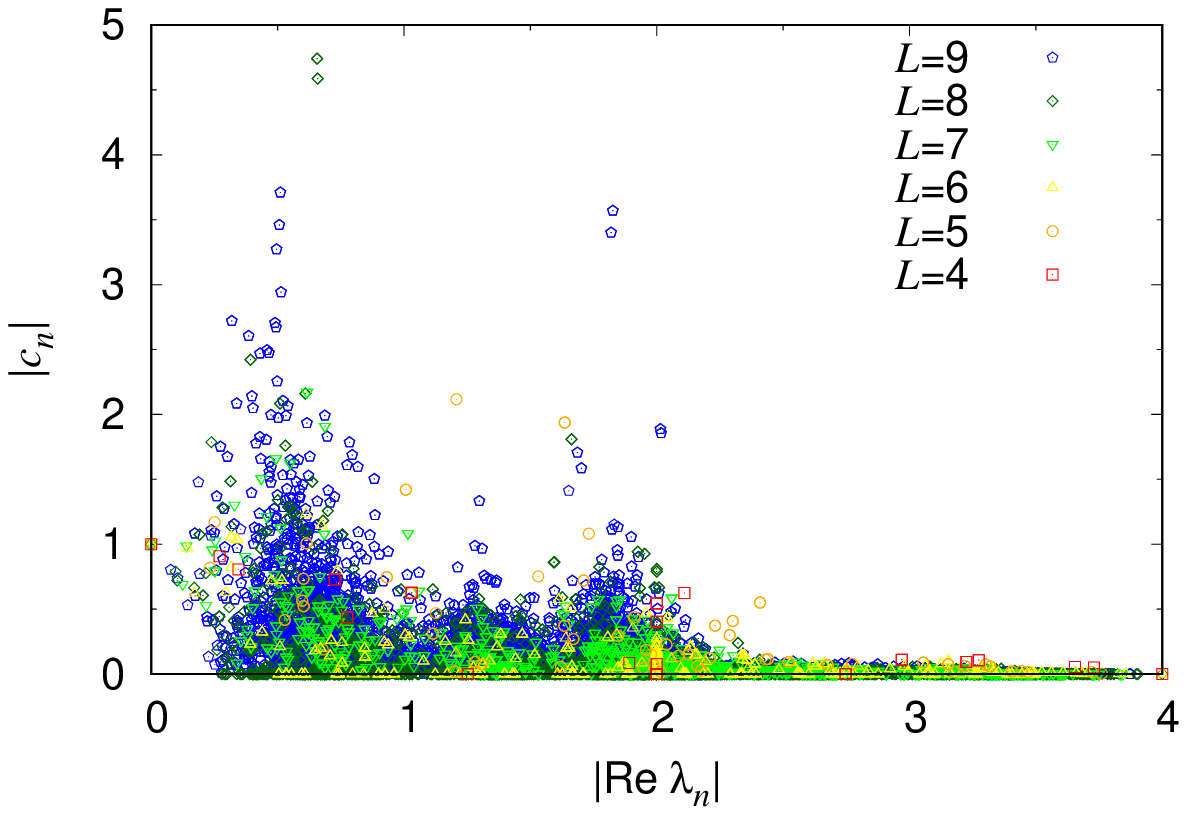}\\
(c) & (d) \\
\includegraphics[width=0.4\linewidth]{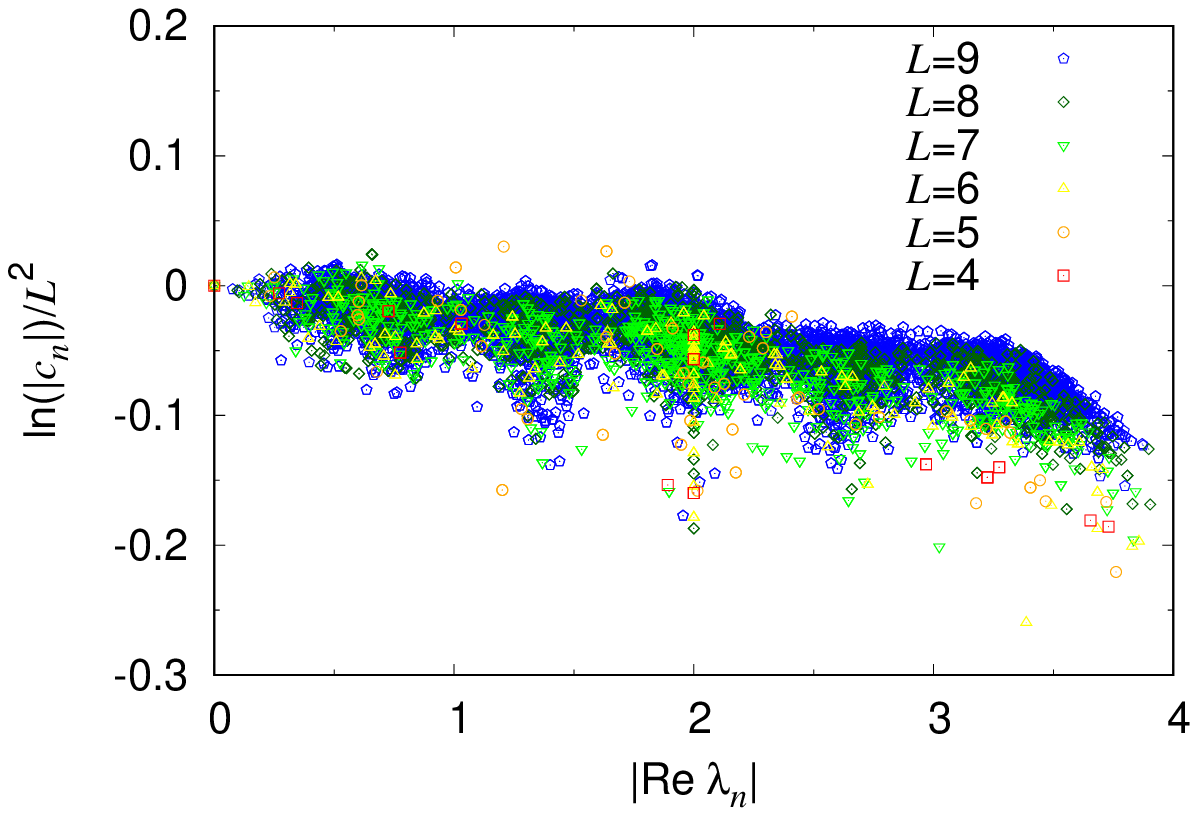}&
\includegraphics[width=0.4\linewidth]{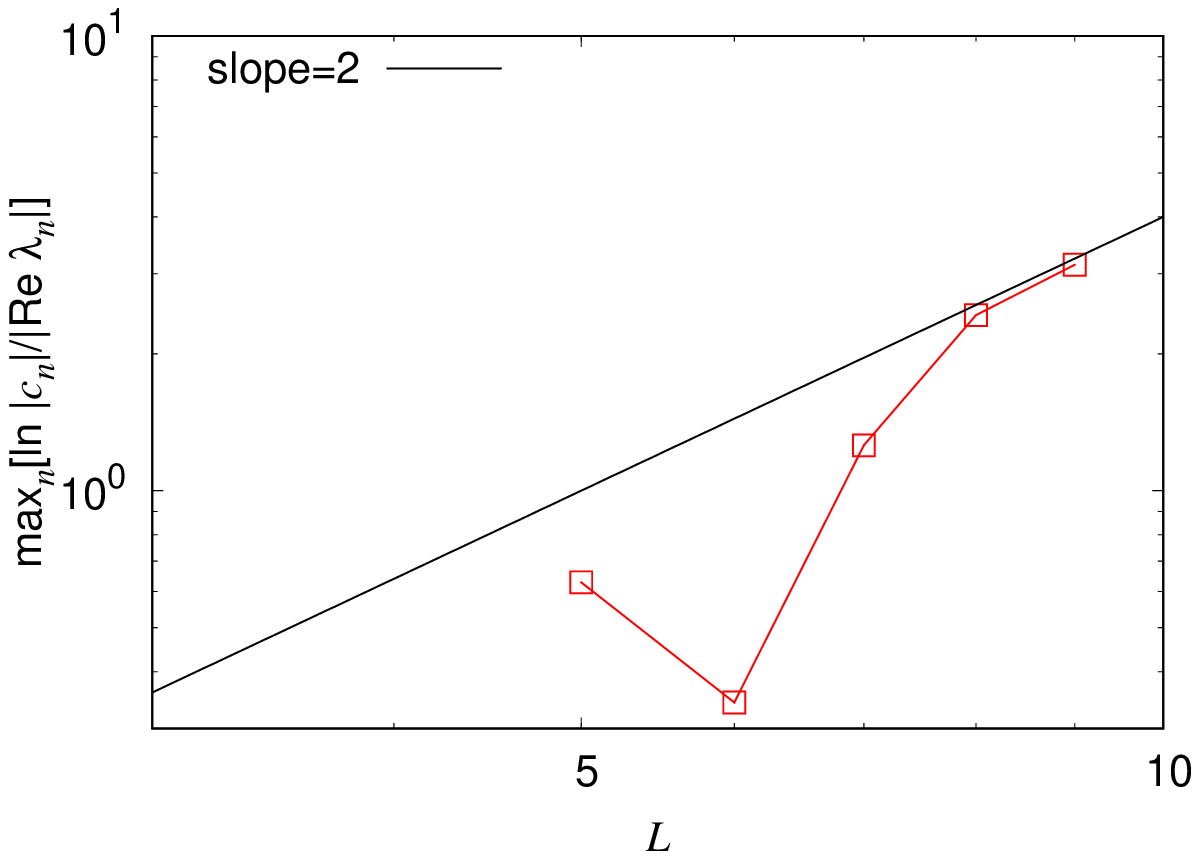}
\end{tabular}
\caption{(a) System-size dependence of $\bar{d}_T(0)=\sum_{n\neq 0}|c_n|$. We find $\ln\bar{d}_T(0)\sim L^2$.
(b) Expansion coefficients $\{|c_n|\}$ against $|\re\lambda_n|$ for the initial state in which all the left-half sites are occupied.
(c) Plot of $\{\ln|c_n|/L^2\}$ against $|\re\lambda_n|$.
(d) Log-log plot of the estimated relaxation time $\tau=\max_n[\ln(|c_n|)/|\re\lambda_n|]$ against $L$.}
\label{fig:expansion}
\end{figure}

\section{B. Other models}
The delay of the relaxation due to large expansion coefficients, which stem from small overlaps between left and right eigenvectors, can generally occur in non-Hermitian dynamics.
The Bose-Hubbard chain under boundary dephasing dissipation is studied in the main text, but here let us present numerical results for other models, i.e., (B.1) the Bose-Hubbard chain under the partcle-driving dissipation, (B.2) the same model under bulk dissipation, and (B.3) the boundary-driven symmetric simple exclusion process (SSEP).

\subsection{B.1 Particle-driving dissipation}

\begin{figure}[b]
\centering
\begin{tabular}{cc}
(a) & (b)\\
\includegraphics[width=0.4\linewidth]{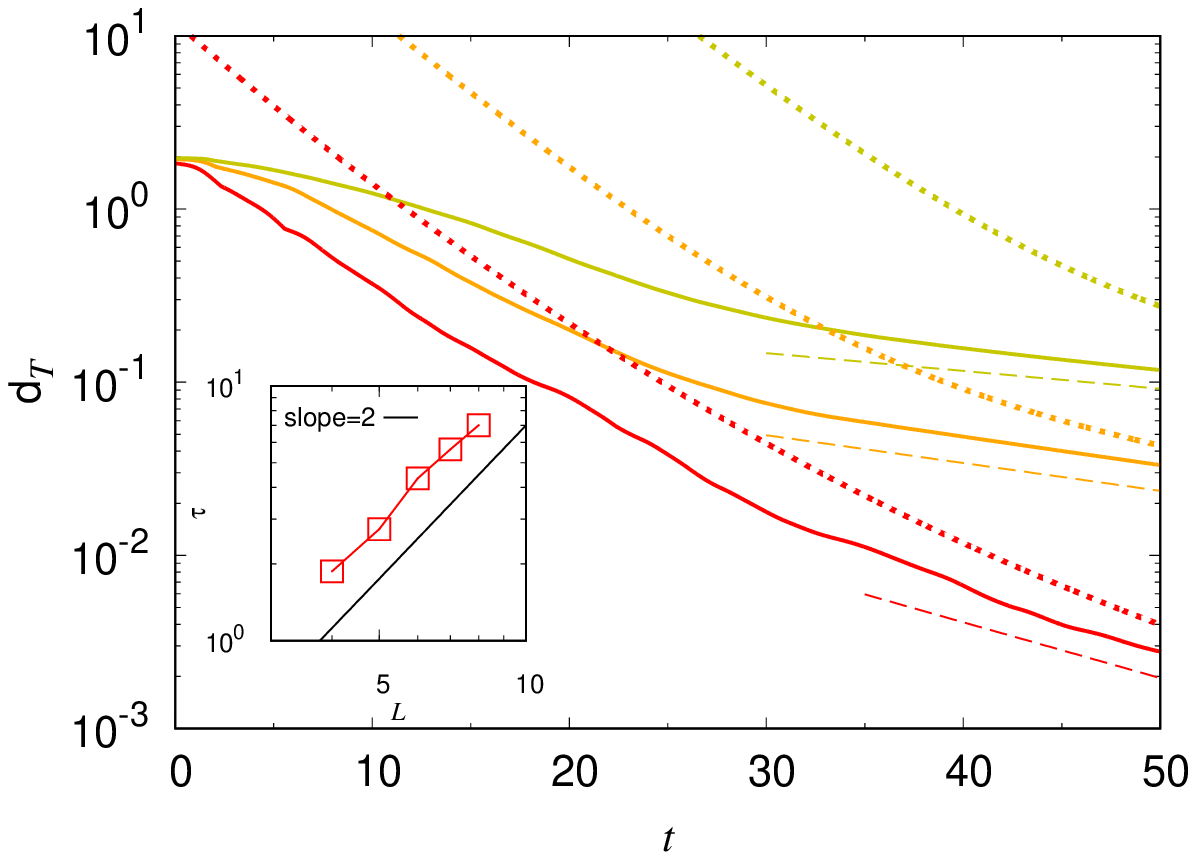}&
\includegraphics[width=0.4\linewidth]{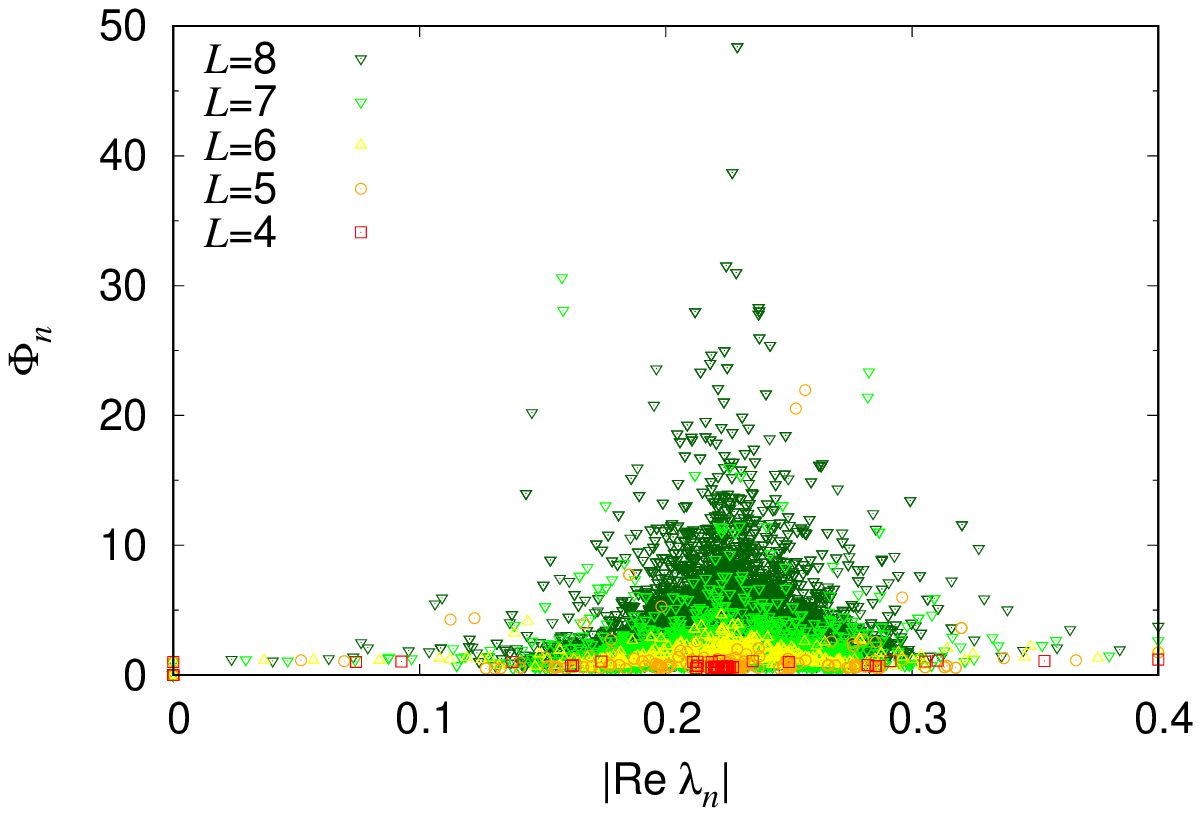}\\
(c) & (d) \\
\includegraphics[width=0.4\linewidth]{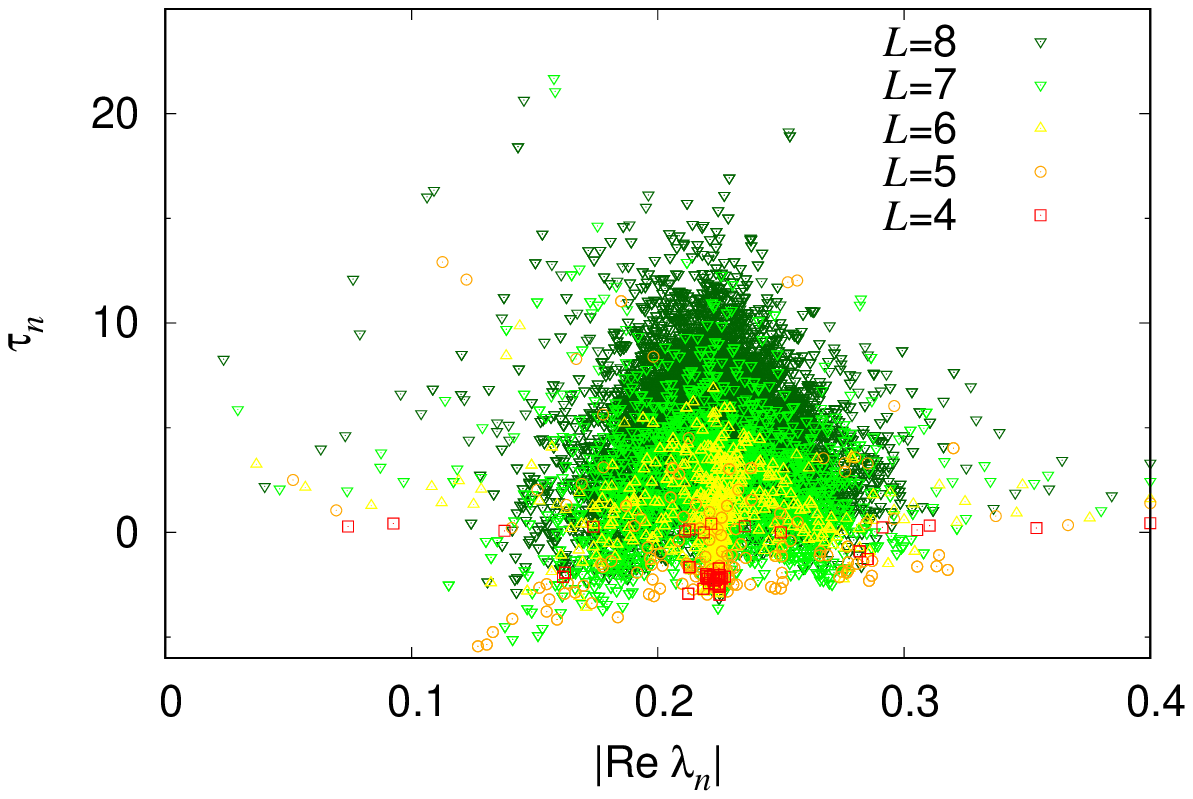}&
\includegraphics[width=0.4\linewidth]{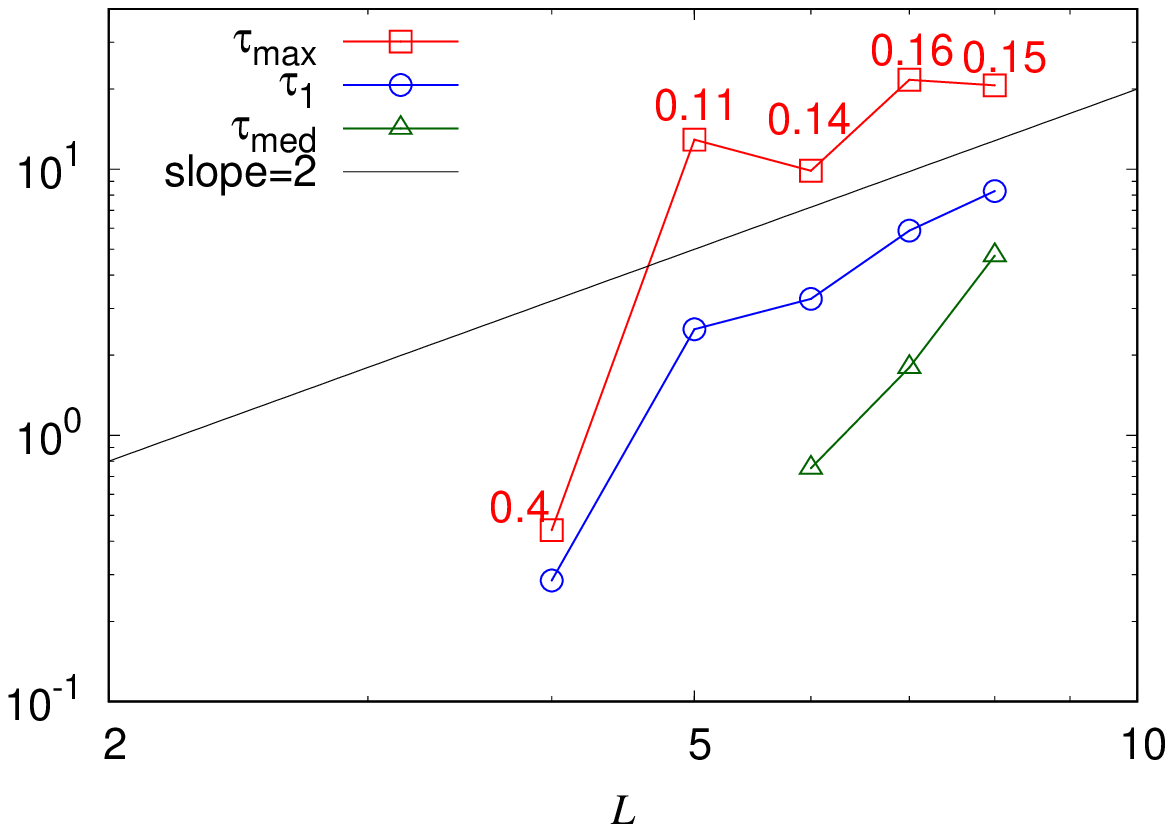}
\end{tabular}
\caption{Numerical results for the system under particle-driving dissipation.
(a) Dynamics of the trace norm distance $d_T(t)$ (solid lines) and its upper bound $\bar{d}_T(t)$ (dotted lines) for $L=4, 6, 8$ from left to right.
Dashed lines show straight lines with the slope $-g$.
The inset shows that the relaxation time $\tau$, which is defined by $d_T(\tau)=1.5$, is proportional to $L^2$.
(b) Numerically calculated values of $\{\Phi_n\}$.
(c) Numerically calculated values of $\{\tau_n\}$.
(d) Log-log plots of $\tau_\mathrm{max}$, $\tau_1$, and $\tau_\mathrm{med}$ against $L$ are shown.
The number indicated for each plot point of $\tau_\mathrm{max}$ in (d) corresponds to $|\re\lambda_{n^*}|$ with $n^*=\argmax_n\tau_n$.}
\label{fig:pd}
\end{figure}

Let us consider the bulk Hamiltonian
\begin{equation}
\hat{H}=-h\sum_{i=1}^{L-1}\left(\hat{b}_{i+1}^\dagger\hat{b}_i+\hat{b}_i^\dagger\hat{b}_{i+1}\right)-h'\sum_{i=1}^{L-2}\left(\hat{b}_{i+2}^\dagger\hat{b}_i+\hat{b}_i^\dagger\hat{b}_{i+2}\right)
\nonumber \\
+U\sum_{i=1}^{L-1}\left(\hat{n}_i-\frac{1}{2}\right)\left(\hat{n}_{i+1}-\frac{1}{2}\right)+U'\sum_{i=1}^{L-2}\left(\hat{n}_i-\frac{1}{2}\right)\left(\hat{n}_{i+2}-\frac{1}{2}\right),
\label{eq:SM_BH}
\end{equation}
with $h=U=1$ and $h'=U'=0.24$, which is the same one studied in the main text.

In addition to the boundary dephasing dissipation, here let us consider boundary dissipation terms which drive the particle flow.
The corresponding Lindblad operators are given by
\begin{equation}
\hat{L}_1=\sqrt{\gamma}\hat{b}_1^\dagger, \quad \hat{L}_2=\sqrt{\gamma'}\hat{b}_1^\dagger\hat{b}_1, \quad \hat{L}_3=\sqrt{\gamma}\hat{b}_L, \quad \hat{L}_4=\sqrt{\gamma'}\hat{b}_L^\dagger\hat{b}_L,
\end{equation}
where newly added Lindblad operators $\hat{L}_1$ and $\hat{L}_3$ represent that a particle is added to the left edge (the site $i=1$) and is removed from the right edge (the site $i=L$) at rate $\gamma$, respectively.
We choose $\gamma=0.2$ and $\gamma'=0.05$.

Let us define a superoperator $\mathcal{N}$ as $\mathcal{N}\rho=[\hat{N},\rho]$, where $\hat{N}=\sum_{i=1}^L\hat{b}_i^\dagger\hat{b}_i$ is the total particle-number operator.
The model under particle-driving dissipation conserves $\mathcal{N}$. If we do not allow superposition of quantum states with different particle numbers, $\mathcal{N}=0$ should hold.
In our numerical calculations, we therefore focus on the sector of $\mathcal{N}=0$.

Our numerical results for $\{\Phi_n\}$ and $\{\tau_n\}$ are shown in Fig.~\ref{fig:pd}.
We find that $\Phi_n=e^{O(L^2)}$ and $\tau_n\propto L^2$ typically hold, which is qualitatively same as in the model under boundary dephasing dissipation.

\subsection{B.2 Bulk dissipation}

\begin{figure}[b]
\centering
\begin{tabular}{cc}
(a) & (b) \\
\includegraphics[width=0.4\linewidth]{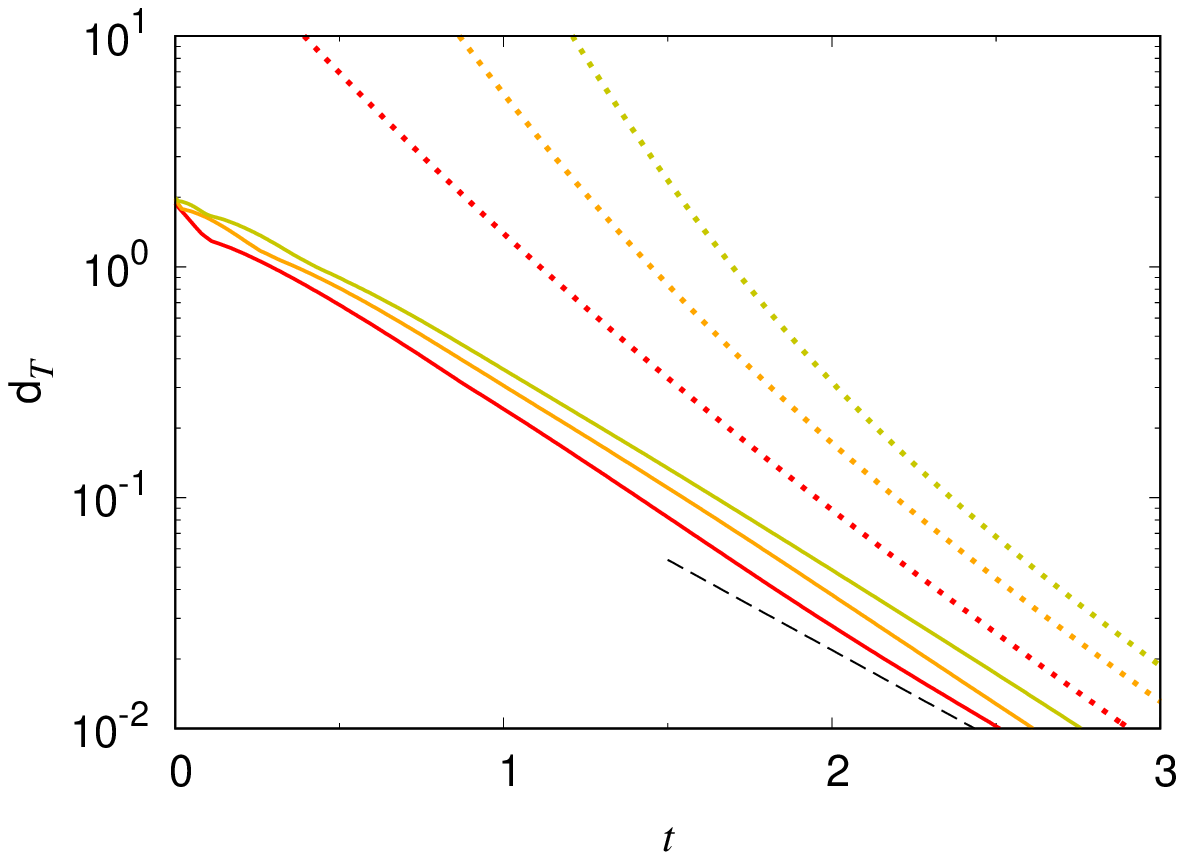}&
\includegraphics[width=0.4\linewidth]{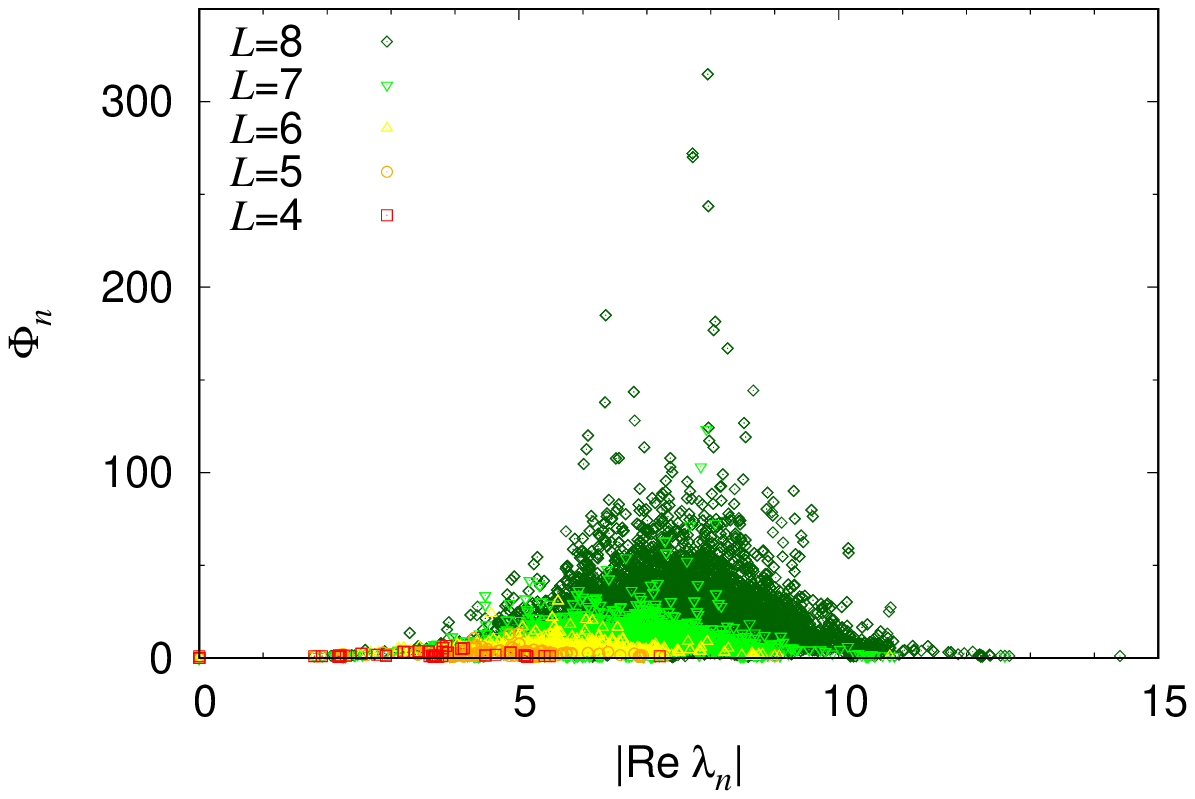}\\
(c) & (d) \\
\includegraphics[width=0.4\linewidth]{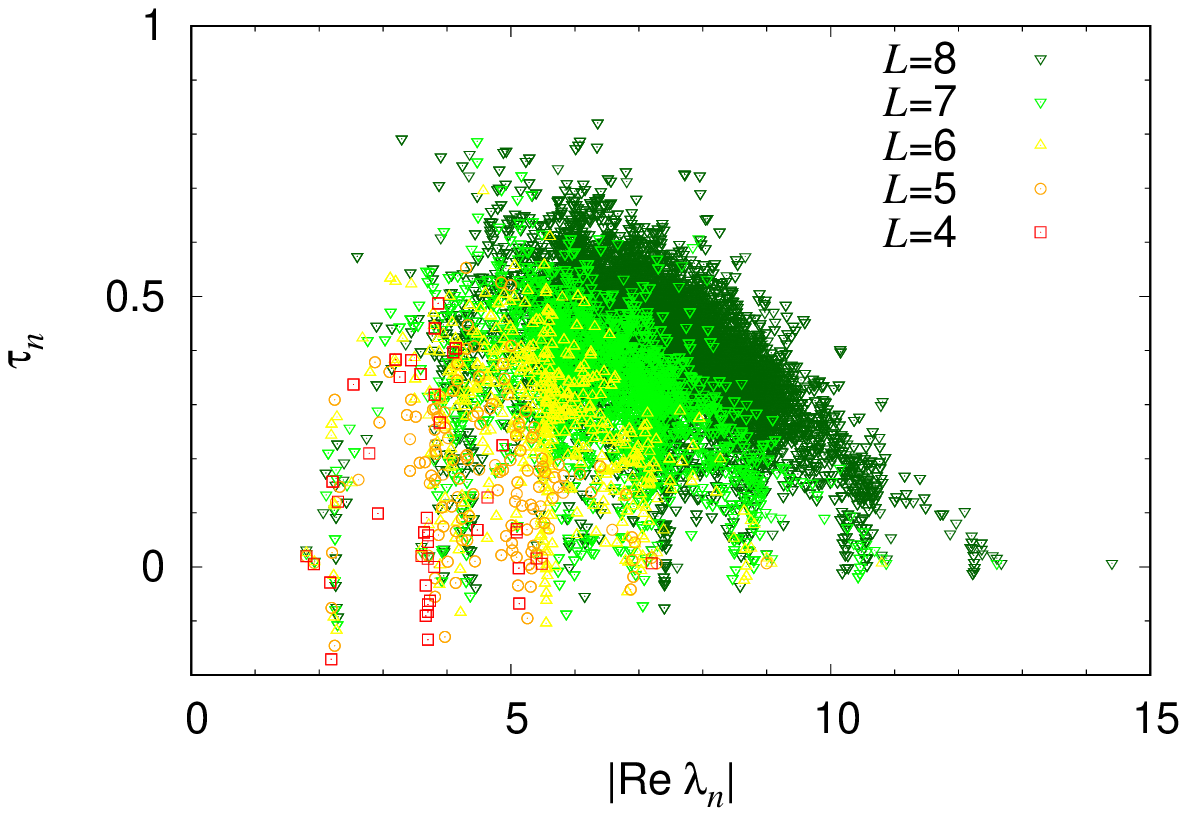}&
\includegraphics[width=0.4\linewidth]{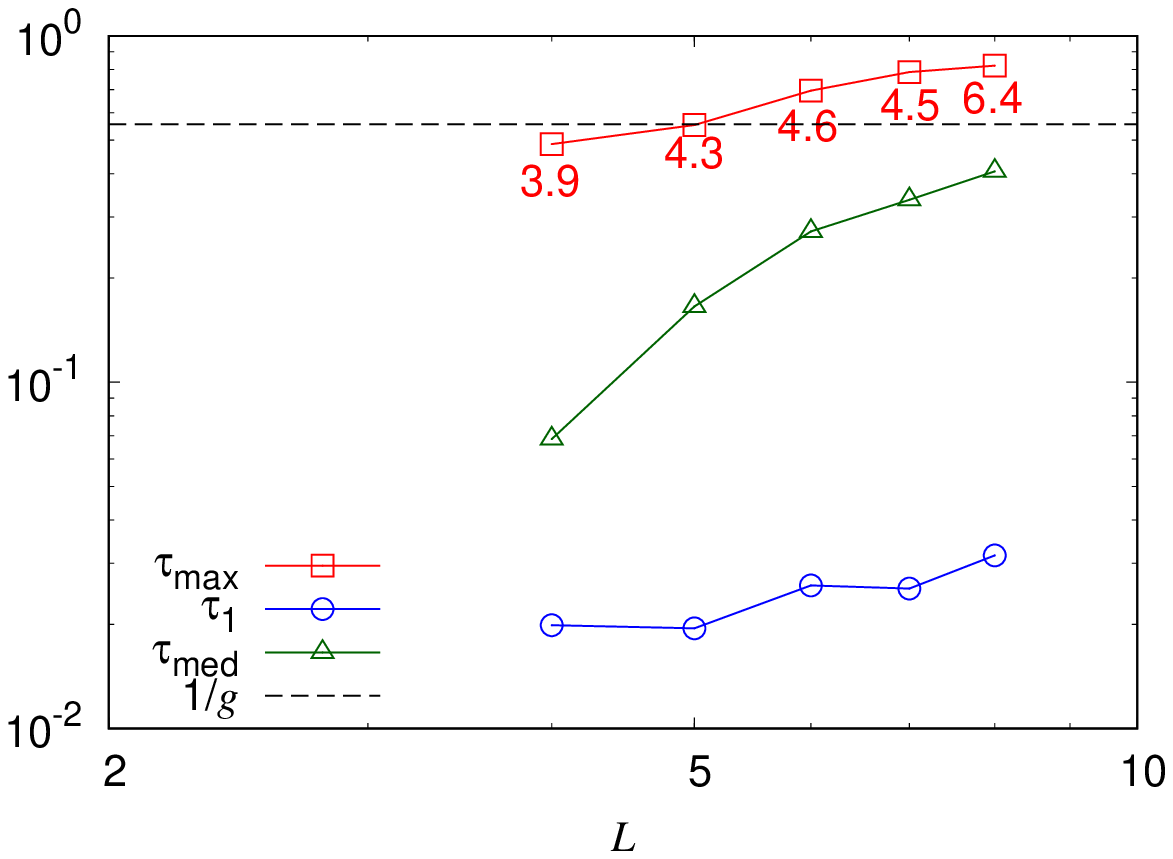}
\end{tabular}
\caption{Numerical results for the system under bulk dissipation.
(a) Dynamics of the trace norm distance $d_T(t)$ (solid lines) and its upper bound $\bar{d}_T(t)$ (dotted lines) for $L=4, 6, 8$ from left to right.
A dashed line shows a straight line with the slope $-g$.
(b) Numerically calculated values of $\{\Phi_n\}$.
(c) Numerically calculated values of $\{\tau_n\}$.
(d) Log-log plots of $\tau_\mathrm{max}$, $\tau_1$, and $\tau_\mathrm{med}$ against $L$ are shown.
The number indicated for each plot point of $\tau_\mathrm{max}$ in (d) corresponds to $|\re\lambda_{n^*}|$ with $n^*=\argmax_n\tau_n$.}
\label{fig:bulk}
\end{figure}

Again we consider the same bulk Hamiltonian~(\ref{eq:SM_BH}) under bulk dissipation.
For each site $i$, we introduce three Lindblad operators 
\begin{equation}
\hat{L}_{i,1}=\sqrt{\gamma_1}\hat{b}_i, \quad \hat{L}_{i,2}=\sqrt{\gamma_2}\hat{b}_i^\dagger, \quad \hat{L}_{i,3}=\sqrt{\gamma_3}\hat{b}_i^\dagger\hat{b}_i
\end{equation}
where we set $\gamma_1=1$, $\gamma_2=0.8$, and $\gamma_3=0.1$.
As in the particle-driving dissipation, we focus on the sector of $\mathcal{N}=0$.

In this model, the gap is almost independent of $L$.
This model does not have any local conserved quantities, and consequently, the relaxation time is finite in the thermodynamic limit.
The system-size dependences of the Liouvillian gap and the relaxation time are thus consistent.

Our numerical results are presented in Fig.~\ref{fig:bulk}.
We see that $\Phi_n$ behaves as $\Phi_n\sim e^{c|\re\lambda_n|}$ for $c>0$ and is peaked at $|\re\lambda_n|=O(L)$ with peak height $e^{O(L)}$.
We do not find superexponential one $\Phi_n=e^{O(L^2)}$.
Correspondingly, $\tau_n=\ln\Phi_n/|\re\lambda_n|$ does not depend on $L$ so much [see Fig.~\ref{fig:bulk} (b) and (c)].
This is an expected result.
However, because of growing $\Phi_n$ with $|\re\lambda_n|$, the maximum of $\tau_n$ comes from a relatively large eigenvalue.
As is clearly seen in Fig.~\ref{fig:bulk} (c), $\tau_1$ is very small, which indicates that the relaxation time and the long-time dynamics are not determined by the first-excited eigenmode.
The dashed line in Fig.~\ref{fig:bulk} (c) represents $g^{-1}$, which is quantitatively comparable to $\tau_\mathrm{max}$ and $\tau_\mathrm{med}$.
However, this quantitative agreement would be accidental, and in principle we need to take the system-size dependence of expansion coefficients into account to accurately evaluate the relaxation time even if there is no obvious discrepancy between the Liouvillian gap and the relaxation time.

\subsection{B.3 Boundary-driven SSEP}

\begin{figure}[b]
\centering
\begin{tabular}{cc}
(a) & (b)\\
\includegraphics[width=0.4\linewidth]{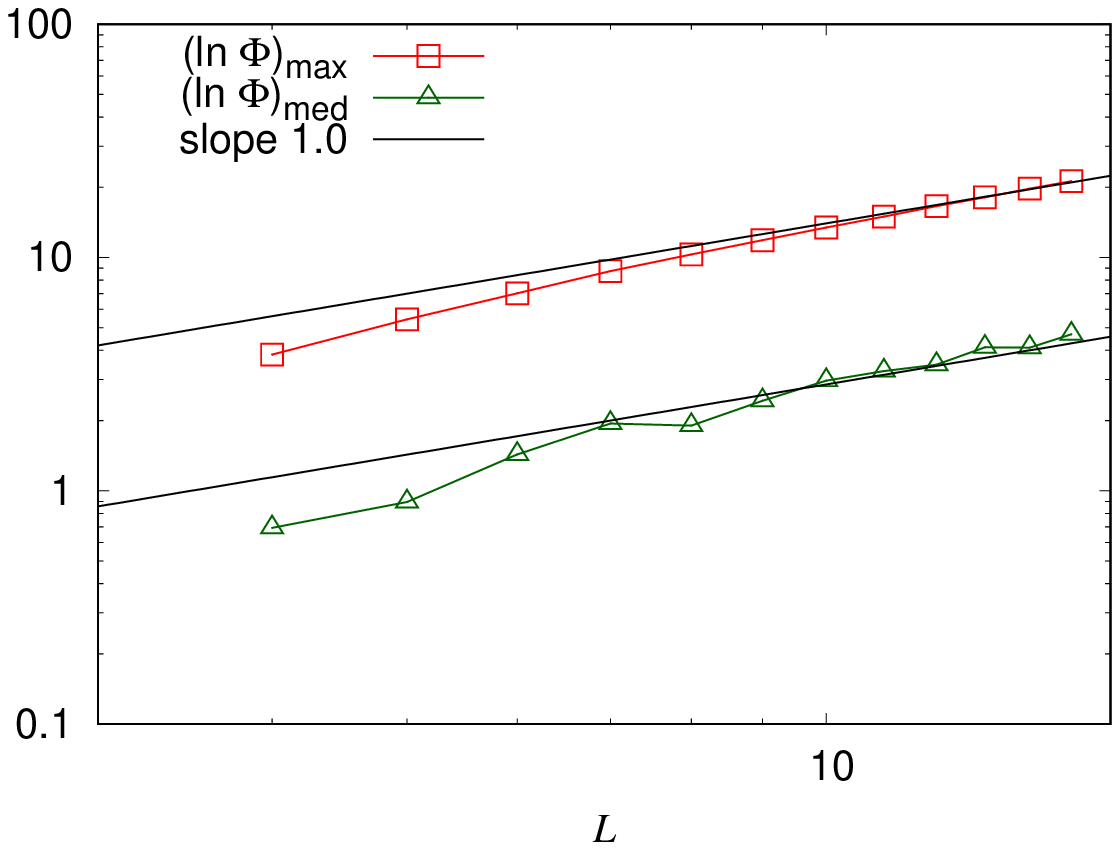}&
\includegraphics[width=0.4\linewidth]{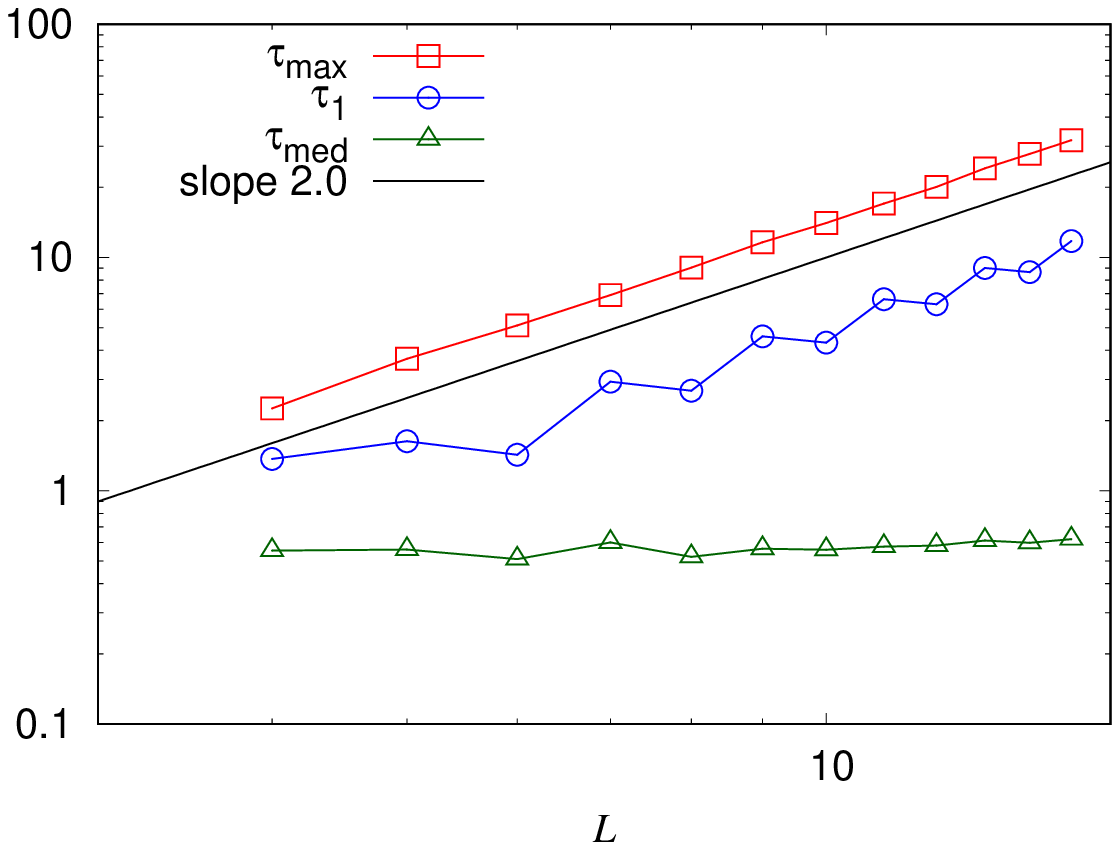}
\end{tabular}
\caption{Log-log plots of (a) $(\ln\Phi)_\mathrm{max}=\max_n\ln\Phi_n$ and $(\ln\Phi)_\mathrm{med}$ denoting the median of $\{\ln\Phi_n\}$, and (b) $\tau_\mathrm{max}$, $\tau_1$, and $\tau_\mathrm{med}$ against the system size $L$.}
\label{fig:SSEP}
\end{figure}

A classical Markov process is also generated by a linear non-Hermitian matrix, and hence the general argument in the main text is also applied.
Here we consider one of the simplest models, i.e., the boundary-driven SSEP.

Let $i=1,2,\dots,L$ be an index of sites, and each site $i$ is either empty ($n_i=0$) or occupied ($n_i=1$) by a single particle.
Each particle attempts to jump to the left or right neighbor at rate 1.
It succeeds only if the target site is empty.
At the left boundary $i=1$, a particle is added at rate $\gamma$ when the site is empty.
At the right boundary $i=L$, a particle is removed at rate $\gamma'$ when the site is occupied.
In this section we choose $\gamma=0.8$ and $\gamma'=0.9$.
The state of the system is specified by $\bm{n}=(n_1,n_2,\dots,n_L)$.
Let us introduce $P_t(\bm{n})$, which is the probability that the system is in the state $\bm{n}$ at time $t$.
The dynamics is described by the classical master equation
\begin{equation}
\frac{\del}{\del t}P_t(\bm{n})=\sum_{\bm{n}'}W(\bm{n},\bm{n}')P_t(\bm{n}'),
\end{equation}
where the $2^L\times 2^L$ matrix $W$ is the transition matrix corresponding to the dynamical rule specified above.
We do not give $W$ explicitly, but instead, we remark that this classical master equation is equivalently expressed by the Lindblad equation for the ``density matrix'' $\rho(t)=\sum_{\bm{n}}P_t(\bm{n})\ket{\bm{n}}\bra{\bm{n}}$.
Using the creation operators $\{b_i^\dagger\}$ and annihilation operator $\{b_i\}$ of hard-core bosons, the state $\ket{\bm{n}}$ is expressed as $\ket{\bm{n}}=(b_{1}^\dagger)^{n_1} (b_{2}^\dagger)^{n_2} \dots (b_{L}^\dagger)^{n_L}\ket{0}$, where $\ket{0}$ is the vacuum.
By introducing the Lindblad operators
\begin{equation}
L_i^r=b_ib_{i+1}^\dagger, \quad L_i^l=b_i^\dagger b_{i+1}, \quad L_1=\sqrt{\gamma}b_1^\dagger, \quad L_L=\sqrt{\gamma'}b_L,
\end{equation}
it turns out that the classical master equation for the boundary-driven SSEP is equivalent to the following Lindblad equation with no Hamiltonian:
\begin{align}
\frac{d}{dt}\rho(t)=\mathcal{L}\rho(t):=\sum_{i=1}^{L-1}\left[\left(L_i^r\rho(t)L_i^{r\dagger}-\frac{1}{2}\{L_i^{r\dagger}L_i^r,\rho(t)\}\right)
+\left(L_i^l\rho(t)L_i^{l\dagger}-\frac{1}{2}\{L_i^{l\dagger}L_i^l,\rho(t)\}\right)\right]
\nonumber \\
+\left(L_1\rho(t)L_1^{\dagger}-\frac{1}{2}\{L_1^{\dagger}L_1,\rho(t)\}\right)
+\left(L_L\rho(t)L_L^{\dagger}-\frac{1}{2}\{L_L^{\dagger}L_L,\rho(t)\}\right)
\end{align}
if we restrict ourselves to the subspace in which every off-diagonal matrix element $\braket{\bm{n}|\rho(t)|\bm{m}} (\bm{n}\neq\bm{m})$ is zero. 

When $\gamma=0$, $W$ (or the Liouvillian $\mathcal{L}$) is Hermitian and $\Phi_n=1$ for all $n$.
While, when $\gamma>0$, the generator is non-Hermitian and $\Phi_n$ may diverge in the thermodynamic limit $L\to\infty$.
Since the particle transport is diffusive in the SSEP, it is expected that the relaxation time is proportional to $L^2$ if the relaxation is associated with the particle diffusion over the entire system.
The Liouvillian gap $g$ in this model is exactly calculated via the Bethe ansatz, according to which it shrinks with $L$ as $g\sim L^{-2}$~\cite{DeGier2005}.
In this case, there is no discrepancy in the system-size dependence of the gap and the relaxation time.

We calculated $\{\Phi_n\}$ and $\{\tau_n\}$ for this model.
It turns out that $\{\Phi_n\}$ typically increase with $L$, but there is no $\Phi_n$ such that $\Phi_n=e^{O(L^2)}$.
According to our numerical result up to $L=15$ in Fig.~\ref{fig:SSEP} (a), both the maximum value and the median of $\ln\Phi_n$ are proportional to $L$.
It means that expansion coefficients are typically exponential in $L$, not $L^2$.
This is a crucial difference from the quantum model discussed in the main text.

On the other hand, if we look at $\tau_\mathrm{max}=\max_n\tau_n$ ($\tau_n=\ln\Phi_n/|\re\lambda_n|$), it is proportional to $L^2$. See Fig.~\ref{fig:SSEP} (b).
The diffusive relaxation time certainly appears, but its mechanism differs from the boundary-dissipated quantum models in the main text.
We find that $\ln\Phi_n\propto L$ for $n$ with $|\re\lambda_n|\sim 1/L$, which leads to $\tau_\mathrm{max}=\max_n\ln\Phi_n/|\re\lambda_n|\propto L^2$ (remember that in the quantum model discussed in the main text, $\ln\Phi_n\propto L^2$ at $|\re\lambda_n|\sim 1$).
Figure~\ref{fig:SSEP} (b) also shows that $\tau_1\propto L^2$, and hence the first excited state that gives the Liouvillian gap is also responsible for the diffusive relaxation time.
In this way, diffusive relaxation in this model is associated with low-lying eigenmodes with eigenvalues $|\re\lambda_n|\lesssim 1/L$ (not necessarily $|\re\lambda_n|\sim 1/L^2$).

Since the bulk dynamics is not unitary, typical values of $\re\lambda_n$ linearly increase with $L$.
On the other hand, $\Phi_n$ typically scales as $\Phi_n=e^{O(L)}$.
As a result, for typical $n$, $\tau_n$ does not depend on $L$; $\tau_n=\ln\Phi_n/|\re\lambda_n|=O(1)$.
This is confirmed by computing the median of $\{\tau_n\}$ for each $L$ [see Fig.~\ref{fig:SSEP} (b)].
This is also different from the quantum model discussed in the main text.

\section{C. Eigenvalue distances and eigenstate co-linearities}

Large expansion coefficients or small overlaps between left and right eigenvectors generally occur when parameters are close to an exceptional point.
At an exceptional point, two or more eigenvalues are degenerate and the corresponding eigenvectors become identical, which makes the matrix not diagonalizable.

As a simple example, let us consider the matrix
\begin{equation}
A=\begin{pmatrix}1&1\\ \varepsilon & 1\end{pmatrix}.
\end{equation}
Obviously $\varepsilon=0$ corresponds to an exceptional point since the matrix is in the Jordan canonical form.
Now we shall consider $\varepsilon>0$.
The eigenvalues are given by $\lambda_{\pm}=1\pm\sqrt{\varepsilon}$, and hence the eigenvalue distance is given by $\delta\lambda=2\sqrt{\varepsilon}$.
The left eigenvector $\vec{\pi}_+$ and the right eigenvector $\vec{\rho}_+$ with the eigenvalue $\lambda_+$ are given by
\begin{equation}
\vec{\pi}_+=\frac{1}{\sqrt{1+\varepsilon}}\begin{pmatrix}\sqrt{\varepsilon} \\ 1\end{pmatrix},
\quad
\vec{\rho}_+=\frac{1}{\sqrt{1+\varepsilon}}\begin{pmatrix}1 \\ \sqrt{\varepsilon}
\end{pmatrix}.
\end{equation}
The inner-product between them is given by
\begin{equation}
\vec{\pi}_+\cdot\vec{\rho}_+=\frac{2\sqrt{\varepsilon}}{1+\varepsilon}\approx\delta\lambda \quad\text{for small }\varepsilon.
\end{equation}
In this way, the overlap between the left and right eigenvector is approximately equal to the eigenvalue distance.
Similarly, the left $\vec{\pi}_-$ and right $\vec{\rho}_-$ eigenvectors of the eigenvalue $\lambda_-$ are given by
\begin{equation}
\vec{\pi}_-=\frac{1}{\sqrt{1+\varepsilon}}\begin{pmatrix}-\sqrt{\varepsilon} \\ 1\end{pmatrix},
\quad
\vec{\rho}_-=\frac{1}{\sqrt{1+\varepsilon}}\begin{pmatrix}1 \\ -\sqrt{\varepsilon}
\end{pmatrix}.
\end{equation}
For small $\varepsilon$, $\vec{\rho}_+$ and $\vec{\rho}_-$, or $\vec{\pi}_+$ and $\vec{\pi}_-$ are almost parallel.
We denote the angle between $\vec{\rho}_+$ and $\vec{\rho}_-$ by $\phi$.
Then we have
\begin{equation}
|\cos\phi|=|\vec{\rho}_+\cdot\vec{\rho}_-|=(1-\varepsilon)/(1+\varepsilon)\approx 1-2\varepsilon=1-\delta\lambda,
\end{equation}
which shows that $\vec{\rho}_+$ and $\vec{\rho}_-$ are almost co-linear.

Let us consider the time evolution of the vector $\vec{v}(t)$ given by $d\vec{v}(t)/dt=-A\vec{v}(t)$.
The initial condition is given by
\begin{equation}
\vec{v}(0)=\begin{pmatrix}0\\ 1\end{pmatrix}.
\end{equation}
This initial state is expanded as
\begin{equation}
\vec{v}(0)=c_+\vec{\rho}_++c_-\vec{\rho}_-=\frac{1}{2}\sqrt{\frac{1+\varepsilon}{\varepsilon}}(\vec{\rho}_+-\vec{\rho}_-).
\end{equation}
We thus have large expansion coefficients $c_+=-c_-\approx 1/\delta\lambda$ for small $\varepsilon$.
However, they do not result in the delay of the relaxation time.
The relaxation time does not diverge as $\varepsilon\to +0$.
In this way, if large expansion coefficients are caused by the near-degeneracy effect, they do not affect the relaxation time.

\begin{figure}[t]
\centering
\includegraphics[width=0.4\linewidth]{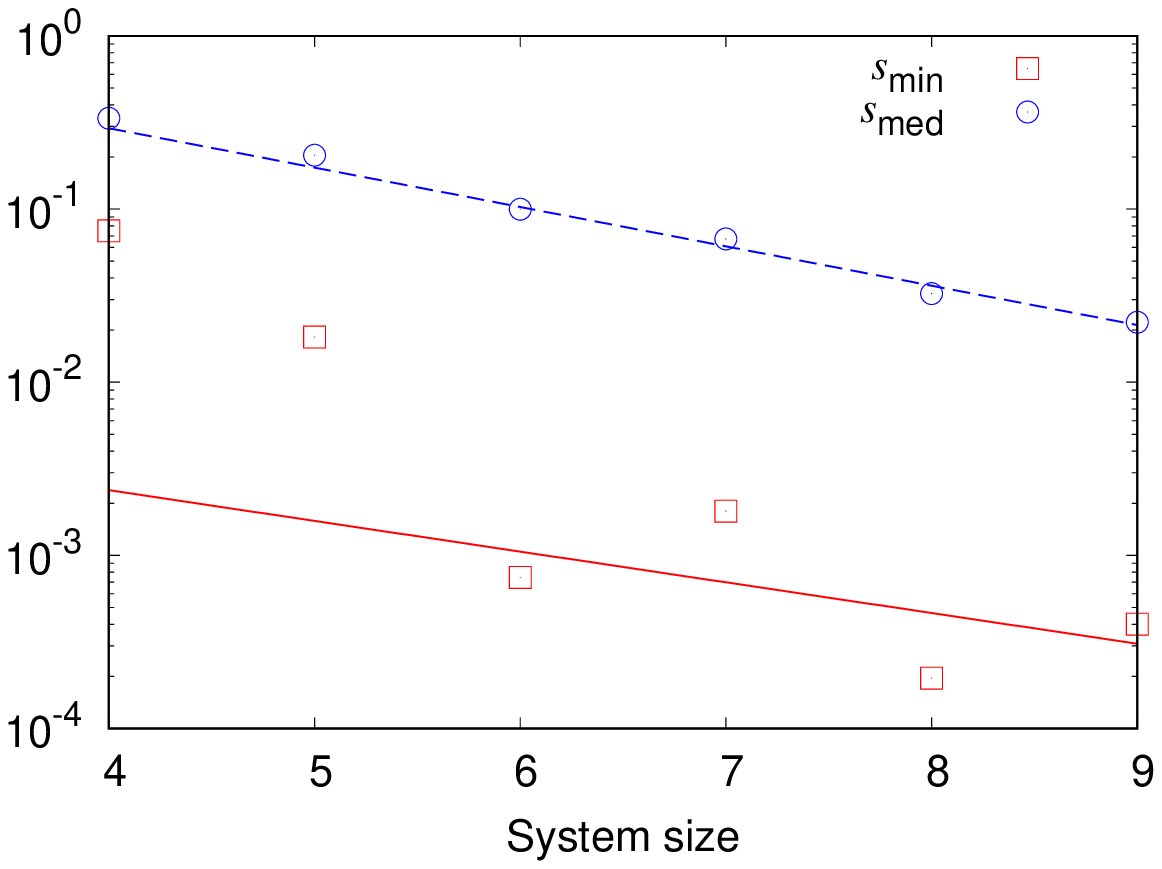}
\caption{System-size dependences of $s_\mathrm{min}$ and $s_\mathrm{med}$. Both of them scale with $L$ as $e^{-O(L)}$.}
\label{fig:level}
\end{figure}

\begin{figure}[t]
\centering
\begin{tabular}{ccc}
(a) $L=5$ & (b) $L=6$ & (c) $L=7$ \\
\includegraphics[width=0.3\linewidth]{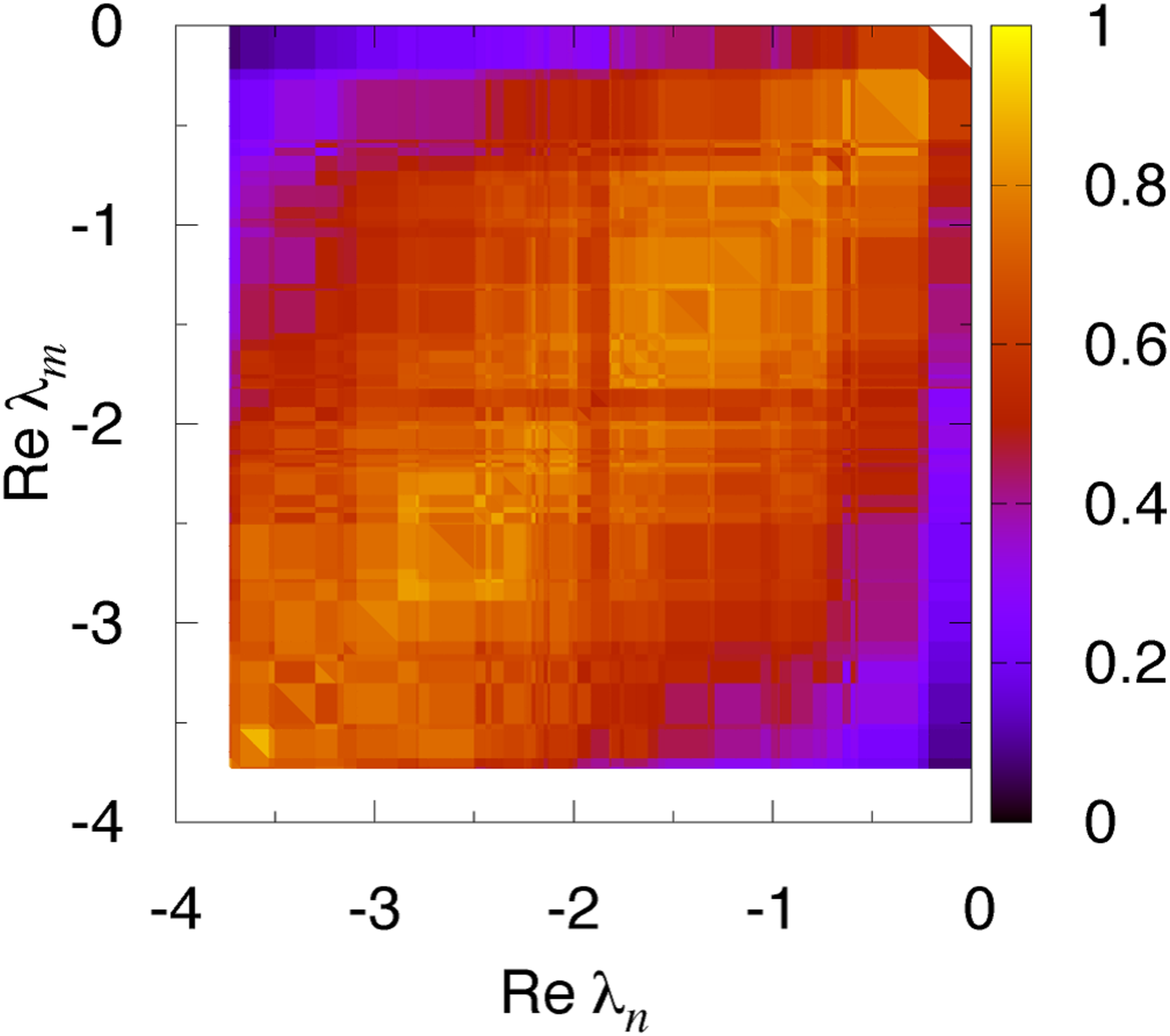}&
\includegraphics[width=0.3\linewidth]{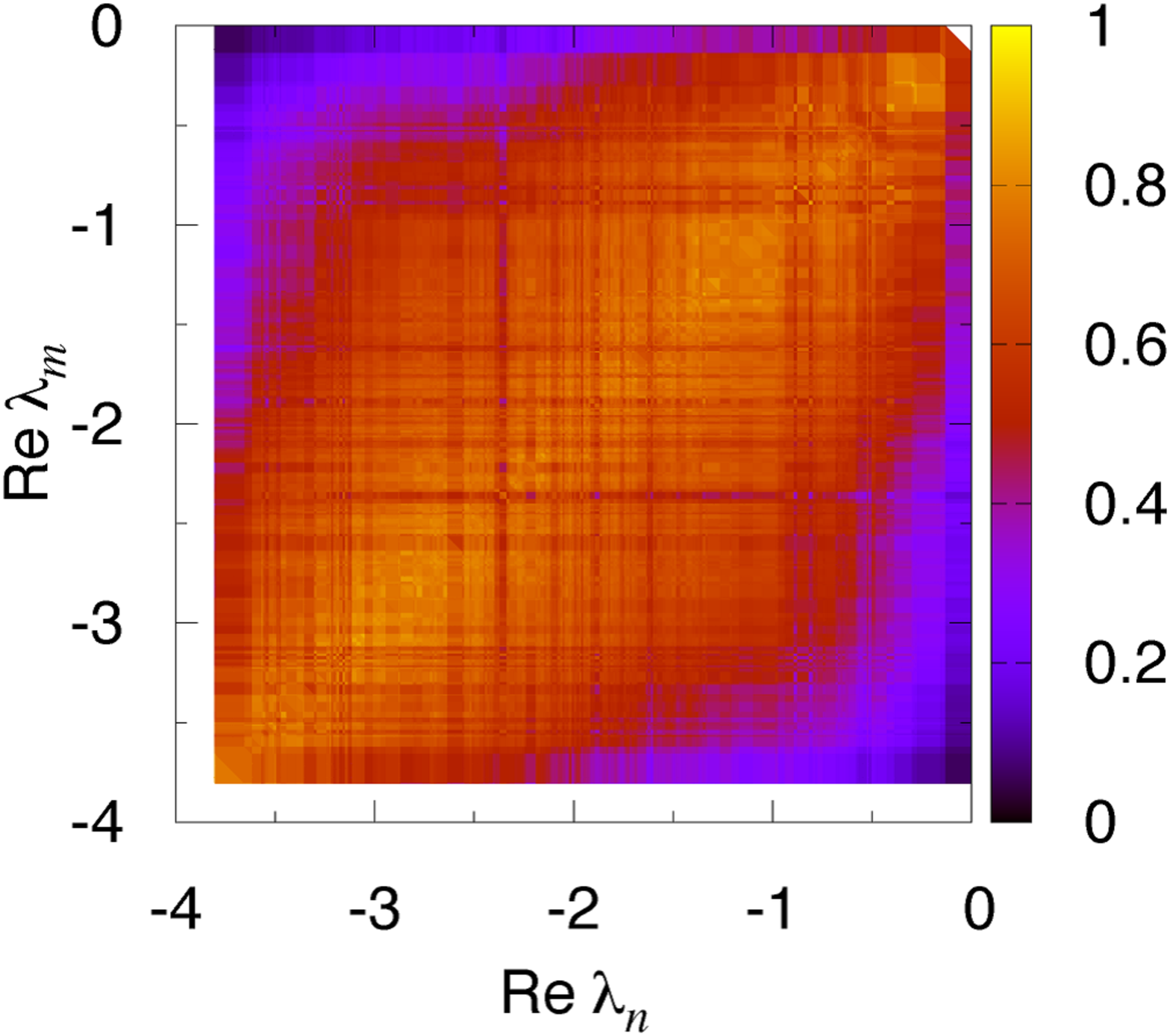}&
\includegraphics[width=0.3\linewidth]{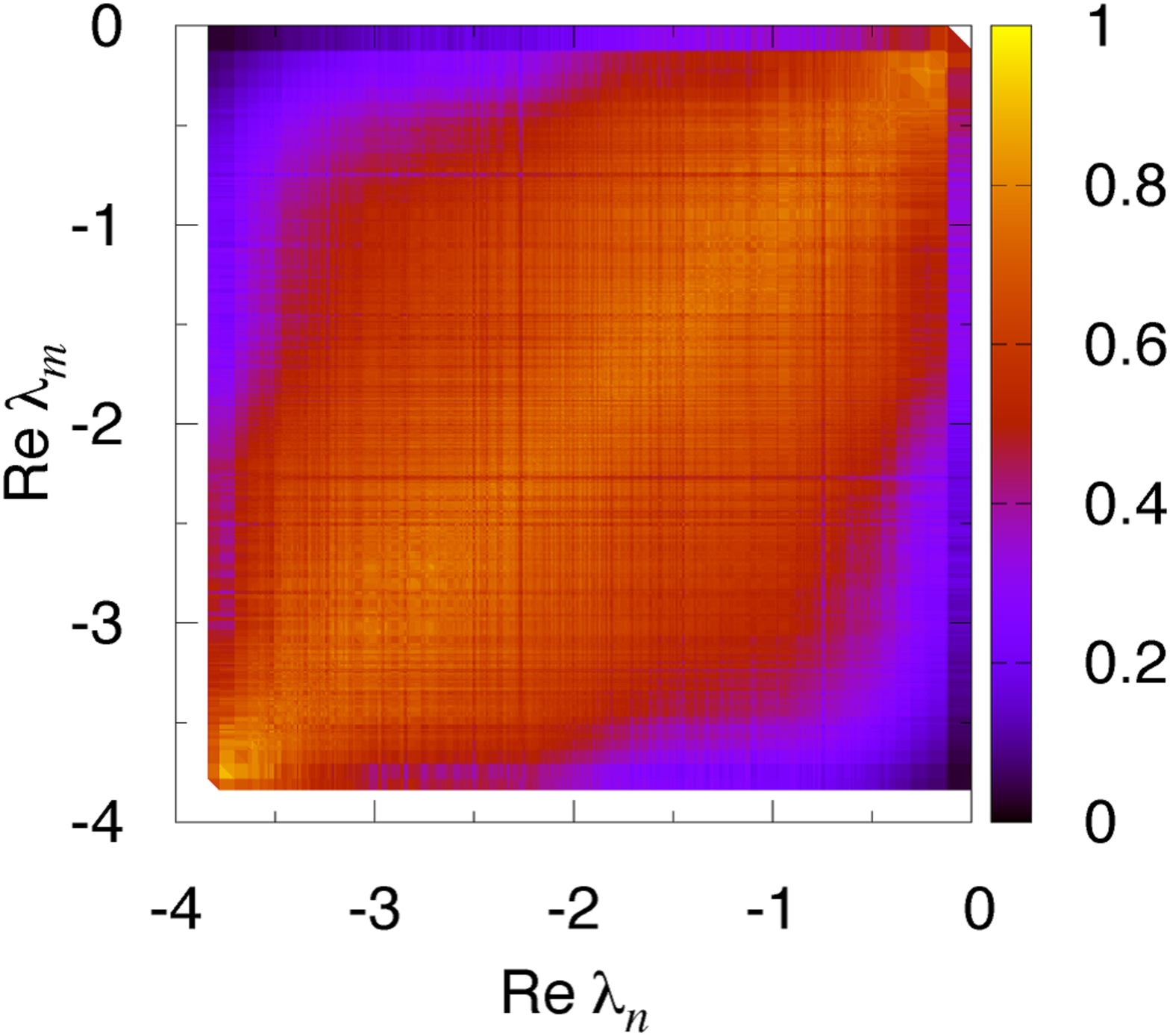}
\end{tabular}
\caption{The values of $p_{nm}$ for all $n\neq m$.
The horizontal and vertical axes are $\re\lambda_n$ and $\re\lambda_m$, respectively.}
\label{fig:p_nm}
\end{figure}

Let us go back to the problem of the boundary-driven open quantum system discussed in the main text.
Now we show that large expansion coefficients $c_n\sim e^{O(L^2)}$, which are explained in the main text, are \textit{not} due to such a trivial near-degeneracy effect.
In order to confirm it, we calculate the eigenvalue distance $s_n$ for each mode $n$ that is defined as
\begin{equation}
s_n=\min_{m(\neq n)}|\lambda_n-\lambda_m|
\end{equation}
and the co-linearity $p_{nm}$ for every $(n,m)$ with $n\neq m$, which is defined as
\begin{equation}
p_{nm}=\frac{|\braket{\rho_n,\rho_m}|}{\sqrt{\braket{\rho_n,\rho_n}\braket{\rho_m,\rho_m}}}.
\end{equation}
When $p_{nm}$ is close to 1, it means that two eigenvectors $\rho_n$ and $\rho_m$ are almost co-linear with each other.
We also denote the minimum distance by $s_\mathrm{min}=\min_n s_n$ and the median value of $\{s_n\}$ by$s_\mathrm{med}$.
Figure~\ref{fig:level} shows the system-size dependence of $s_\mathrm{min}$ and $s_\mathrm{med}$.
We see that both of them scale as $e^{-O(L)}$, which is extremely larger than the typical overlap $\braket{\pi_n,\rho_n}=e^{-O(L^2)}$.
If the small overlap stems from two almost-degenerate eigenvectors, the overlap should be of the same order as their eigenvalue distance.
Our numerical result in Fig.~\ref{fig:level} shows that it is not the case; extremely small overlaps (or extremely large expansion coefficients) are not understood by the trivial near-degeneracy effect, and they can affect the relaxation time.

This conclusion is strengthened by looking at the profile of $p_{nm}$, which is shown in Fig.~\ref{fig:p_nm}.
We see that two eigenmodes with similar eigenvalues are more parallel than those with largely different eigenvalues, but there is no pair of $n\neq m$ such that $p_{nm}\approx 1$ in contrary to the simple case of a $2\times 2$ matrix $A$ near an exceptional point.
In this way, extremely small overlap $\braket{\pi_n,\rho_n}$ is not understood as a trivial effect near an exceptional point.

It would be interesting to see that $p_{nm}$ looks typically $O(1)$, which is much larger than the value expected for the case in which eigenvectors point to independent random directions (in this case we expect $p_{nm}\sim 1/D^2=e^{-O(L)}$, where $D$ is the dimension of the Hilbert space).

\clearpage
\end{widetext}
\end{document}